\documentclass[a4paper,11pt]{article}
\usepackage[utf8]{inputenc}
\usepackage[english]{babel}
\usepackage[T1]{fontenc}
\usepackage{amsbsy, amsfonts, amsmath, amssymb, amsthm}
\usepackage[pdftex]{color}
\usepackage{enumerate, float, calc, color}
\usepackage{multirow}
\usepackage{url} 
\usepackage{hyperref}
\usepackage[makeroom]{cancel}
\usepackage{xcolor}
\usepackage{graphicx}
\usepackage{mathtools}
\usepackage{booktabs}
\usepackage{multirow}
\usepackage{caption}
\usepackage{subcaption}
\usepackage[sectionbib]{natbib}
\usepackage{amsmath}
\usepackage[linesnumbered,ruled,vlined]{algorithm2e}

\newtheorem{theorem}{Theorem}[section]
\newtheorem{definition}{Definition}[section]

\newtheorem{proposition}[theorem]{Proposition}

\newtheorem{remark}[theorem]{Remark}

\usepackage[colorinlistoftodos,textwidth=1.8cm, textsize=tiny]{todonotes}

\hypersetup{
  colorlinks=true,
  linkcolor=blue,
  citecolor=blue,
  filecolor=blue,
  urlcolor=blue,
}

\usepackage[
top    = 2cm,
bottom = 2.50cm,
left   = 1.5cm,
right  = 2.0cm]{geometry}

\usepackage{setspace} 
\begin{document}

 \title{\bf Multivariate Conway-Maxwell-Poisson Distribution: Sarmanov Method and Doubly-Intractable Bayesian Inference}
  	\author{
  	Luiza S.C. Piancastelli$^{1,}$\footnote{Email: \textcolor{teal}{\texttt{luiza.piancastelli@ucdconnect.ie}}}\,,\,\, Nial Friel$^{1, 2,}$\footnote{Email: \textcolor{teal}{\texttt{nial.friel@ucd.ie}}}\,,\,\,
  Wagner Barreto-Souza$^{3,}$\footnote{Email: \textcolor{teal}{\texttt{wagner.barretosouza@kaust.edu.sa}}}\,\, and Hernando Ombao$^{3,}$\footnote{Email: \textcolor{teal}{\texttt{hernando.ombao@kaust.edu.sa}}}\hspace{.2cm}\\
  		{\normalsize \it $^1$School of Mathematics and Statistics, University College Dublin, Dublin, Ireland}\\
  		{\normalsize \it $^2$Insight Centre for Data Analytics, Ireland}\\
  		{\normalsize \it $^3$Statistics Program, King Abdullah University of Science and Technology, Thuwal, Saudi Arabia}}
  	\maketitle

\begin{abstract}
In this paper, a multivariate count distribution with Conway-Maxwell (COM)-Poisson marginals is proposed. To do this, we develop a modification of the Sarmanov method for constructing multivariate distributions. Our multivariate COM-Poisson (MultCOMP) model has desirable features such as (i) it admits a flexible covariance matrix allowing for both negative and positive non-diagonal entries; (ii) it overcomes the limitation of the existing bivariate COM-Poisson distributions in the literature that do not have COM-Poisson marginals; (iii) it allows for the analysis of multivariate counts and is not just limited to bivariate counts. 
Inferential challenges are presented by the likelihood specification as it depends on a number of intractable normalizing constants involving the model parameters. These obstacles motivate us to propose a Bayesian inferential approach where the resulting doubly-intractable posterior is dealt with via the exchange algorithm and the Grouped Independence Metropolis-Hastings algorithm. Numerical experiments based on simulations are presented to illustrate the proposed Bayesian approach. We analyze the potential of the MultCOMP model through a real data application on the numbers of goals scored by the home and away teams in the Premier League from 2018 to 2021. Here, our interest is to assess the effect of a lack of crowds during the COVID-19 pandemic on the well-known home team advantage. A MultCOMP model fit shows that there is evidence of a decreased number of goals scored by the home team, not accompanied by a reduced score from the opponent. Hence, our analysis suggests a smaller home team advantage in the absence of crowds, which agrees with the opinion of several football experts.
\end{abstract}


\medskip


\section{Introduction}

The Conway-Maxwell-Poisson (COM-Poisson) distribution was introduced by \cite{com1962} in the context of queuing systems and was later revived in the literature by \cite{shmetal2005}. A random variable $X$ follows a COM-Poisson distribution if its probability function can be written as
\begin{eqnarray}\label{unicp_pmf}
p(x|\lambda, \nu)=\frac{\lambda^x}{(x!)^\nu Z(\lambda,\nu)}, \quad x\in\mathbb N_0\equiv\{0,1,2,\ldots\},	
\end{eqnarray}
where 
\begin{eqnarray}\label{norm_const}
Z(\lambda,\nu)\equiv\displaystyle\sum_{x=0}^\infty\dfrac{\lambda^x}{(x!)^\nu}
\end{eqnarray}
is the normalised constant, with $\lambda>0$ and $\nu>0$, the latter of which is responsible for controlling the dispersion. 
The COM-Poisson distribution is overdispersed when $\nu< 1$ and underdispersed when $\nu>1$. The case $\nu=1$ corresponds to the Poisson distribution, which is equidispersed. The Bernoulli and geometric distributions can also be obtained from the COM-Poisson. The geometric special case corresponds to $\nu=0$ and $\lambda<1$ while the Bernoulli($\lambda/(\lambda+1)$) distribution is a limiting case as $\nu \rightarrow \infty$.
This distribution has received much attention in the literature after its revival in 2005. Properties of the COM-Poisson distribution are discussed for instance in \cite{nad2009} and \cite{dalgau2016}, and a regression model is introduced by \cite{selshm2010}. Bayesian inference approaches for this model were proposed by \cite{kadetal2006} and \cite{benfri2020}. Approximations for the intractable normalizing constant $Z(\lambda,\nu)$ given in (\ref{norm_const}) have been studied by \cite{dalgau2016} and \cite{gauetal2019}.  Recent contributions on time series analysis and tree-based semi-varying coefficient model are due to \cite{seletal2020} and \cite{chashm2020}, respectively. For a recent account on the COM-Poisson model, we refer the reader to \cite{selpre2020}.

A natural point of interest is the proposal of a multivariate count model with COM-Poisson marginals. A first bivariate proposal attempt in this direction was addressed by \cite{seletal2016} but the marginals of that proposed bivariate model are not longer COM-Poisson distributed. Moreover, the dispersion parameter is assumed the same for both marginals and the range of correlation depends on it, which limits its ability to account dependency. For example, for the particular case for their model when the dispersion equals 1 (the bivariate Poisson case), the correlation is non-negative. The quantities involved in such a bivariate distribution are also very cumbersome; for instance, see the joint probability function given in equation (16) of \cite{seletal2016} which depends on an infinite summation.

Here we aim to address these issues and through the construction of a multivariate COM-Poisson distribution. To do this, we develop a modification of the \cite{sar1966} method for constructing multivariate distributions.  Our multivariate COM-Poisson (MultCOMP) model has desirable features such as (i) flexible covariance matrix allowing for both negative and positive non-diagonal entries; (ii) it overcomes the limitation of the existing bivariate COM-Poisson distribution of \cite{seletal2016} which has neither COM-Poisson marginals nor assumes different dispersion parameters for the marginals; (iii) it allows for analysis of multivariate counts rather than being limited to bivariate counts. A challenging point that arises in our proposed multivariate COM-Poisson model is that the likelihood function depends on the ratio of normalised constants arising from  (\ref{norm_const}). We propose a Bayesian inference based on doubly-intractable posterior via the exchange algorithm and Grouped Independence Metropolis-Hastings to deal with this posed challenge. A recent related work is due to \cite{ongetal2021} where a bivariate COM-Poisson distribution is proposed based on the Sarmanov method. The exponential kernel case discussed there is a particular case of our MultCOMP model when the dimensional equals 2. Furthermore, the inference in that paper is performed via direct maximization of the log-likelihood function without exploring the difficulties involving the appearance of the constants due to (\ref{norm_const}) and the parameter restrictions to be considered in the optimization. Such challenging points are carefully addressed in our paper under a Bayesian perspective.

The remainder of this paper is organized as follows. In Section \ref{sarmanov}, the Sarmanov construction of bivariate distributions is reviewed along with its multivariate extension by \cite{lee1996}. A variation of the former with a lower number of parameters and more tractable correlation bounds is developed in this paper and related properties are discussed. We propose a multivariate COM-Poisson distribution using our modified Sarmanov method in Section \ref{mcomp}. We develop and compare Bayesian methods to deal with intractability of the proposed model likelihood in Section \ref{intractable_terms}. Two MCMC strategies are developed in Section \ref{inference} and compared via simulation studies in Section \ref{simulations}. In Section \ref{data_application}, we apply the MultCOMP model to analyze the numbers of goals scored by the home and away teams in the Premier League from 2018 to 2021. Here, our interest is to assess the effect of the absence of crowds due to the COVID-19 pandemic on the well-known home team advantage. The analysis using the MultCOMP model reveals that the home team advantage, in the Premier League, was significantly diminished while no crowds were allowed in the games.
Concluding remarks are given in Section \ref{conclusions}.


\section{Generalized Sarmanov method}\label{sarmanov}

\cite{sar1966} proposed a method for constructing bivariate distributions with given marginals. This method was extended by \cite{lee1996} in order to accommodate higher-order dimensions rather than two-dimensional. In this section, we propose a modification of the version by \cite{lee1996}, that is more mathematically tractable as explained in what follows. We begin by briefly exploring the works by \cite{sar1966} and \cite{lee1996}.

Let $f_1$ and $f_2$ be two density functions with respect to measures $\mu_1$ and $\mu_2$ with support on $\mathcal S_1$ and $\mathcal S_2$, respectively. Commonly, these are either the counting and Lebesgue measures corresponding to the discrete and continuous cases, respectively. 
A joint density function $f(\cdot,\cdot)$ with respect to the product measure $\mu_1\times\mu_2$ having marginals $f_1$ and $f_2$ is now constructed using the Sarmanov method by
\begin{eqnarray}\label{jointdensity}
f(x_1,x_2)=f_1(x_1)f_2(x_2)\left\{1+\delta\phi_1(x_1)\phi_2(x_2)\right\}, \quad x_1\in\mathcal S_1,  x_2\in\mathcal S_2,
\end{eqnarray}
where $\phi_1(\cdot)$ and $\phi_2(\cdot)$ are bounded functions satisfying $\int_{\mathcal S_i}f_i(x)\phi_i(x)d\mu_i(x)=0$, for $i=1,2$, with $\mathcal S_1$ and $\in\mathcal S_2$ being the marginal supports. Furthermore, these functions and $\delta$ must satisfy $1+\delta\phi_1(x_1)\phi_1(x_2)\geq0$ $\forall(x_1,x_2)\in\mathcal S_1\times \mathcal S_2$ to ensure that (\ref{jointdensity}) is a proper joint density function. As discussed in \cite{ver2020}, the function $\phi_i(\cdot)$ usually assumes the form $\phi_i(x)=u_i(x)-\Psi_i$, where $\Psi_i=E(u_i(X_i))$, with $X_i$ being a random variable having density function $f_i$, for $i=1,2$. Some possible choices for the function $u_i(\cdot)$ are: (i) $u_i(x)=e^{-\omega x}$, which is known as exponential kernel and will be the focus of our paper; (ii) $u_i(x)=x^\omega$ (assuming that the associated support is compact and that $u_i(X_i)$ is integrable); and (iii) $u_i(x)=f_i(x)$. For more details on the bivariate Sarmanov distributions; see \cite{kotetal2000} and \cite{ver2020}.

Let us consider the exponential kernel case, that is $u_i(x)=e^{-\omega x}$ with either $\mathcal S_i=\mathbb N_0$ or $\mathcal S_i=\mathbb R^+$ for $i=1,2$. As mentioned above, in order to ensure that (\ref{jointdensity}) is a proper density function, it is necessary that the following condition holds:
\begin{eqnarray}\label{condition}
1+\delta(e^{-\omega x_1}-\Psi_1)(e^{-\omega x_2}-\Psi_2)>0\quad \forall x_1,x_2\geq0.
\end{eqnarray}
The range of the $\delta$ parameter yielding a valid joint density function for the above case was studied for instance in \cite{lee1996}. From Corollary 2 of that paper, we obtain that the parameter space of $\delta$ ensuring (\ref{condition}) is given by $$-\dfrac{1}{\max\{(1-\Psi_1)(1-\Psi_2),\Psi_1\Psi_2\}}<\delta<\dfrac{1}{\max\{\Psi_1(1-\Psi_2),\Psi_2(1-\Psi_1)\}}.$$ 

Let $(X_1,X_2)$ be a bivariate vector following a bivariate Sarmanov distributions with the exponential kernel function as described above. Denote $\mu_1\equiv E(X_1)$, $\mu_2\equiv E(X_2)$, $\sigma_1\equiv \sqrt{\mbox{Var}(X_1)}$ and $\sigma_2\equiv \sqrt{\mbox{Var}(X_2)}$. The correlation between $X_1$ and $X_2$ is
\begin{eqnarray}\label{corr}
\mbox{corr}(X_1,X_2)=\dfrac{\delta}{\sigma_1\sigma_2}\{\Psi_1'\Psi_2'-\mu_1\Psi_1\Psi_2'-\mu_2\Psi_1'\Psi_2+\mu_1\mu_2\Psi_1\Psi_2\},
\end{eqnarray}
where $\Psi_i'=M_{X_i}'(-\omega)$ for $i=1,2$, with $M_X'(\cdot)$ denoting the first derivative of the marginal moment generation function of a random variable $X$. With the above results, we obtain that the range of correlation between $X_1$ and $X_2$ is
\begin{eqnarray}\label{rangecorbiv}
	-\dfrac{\Psi_1'\Psi_2'-\mu_1\Psi_1\Psi_2'-\mu_2\Psi_1'\Psi_2+\mu_1\mu_2\Psi_1\Psi_2}{\sigma_1\sigma_2\max\{(1-\Psi_1)(1-\Psi_2),\Psi_1\Psi_2\}}<\mbox{corr}(X_1,X_2)<\dfrac{\Psi_1'\Psi_2'-\mu_1\Psi_1\Psi_2'-\mu_2\Psi_1'\Psi_2+\mu_1\mu_2\Psi_1\Psi_2}{\sigma_1\sigma_2\max\{\Psi_1(1-\Psi_2),\Psi_2(1-\Psi_1)\}}.
\end{eqnarray} 

The Sarmanov method was extended by \cite{lee1996} in order to allow the construction of higher-order multivariate distributions rather than bivariate; see also \cite{kotetal2000}. Let $f_1,\ldots,f_d$ be $d\in\mathbb N$ density functions with respect to the measures $\mu_1,\ldots,\mu_d$ with respective supports $\mathcal S_1,\ldots,\mathcal S_d$. Then, a joint density function having marginal densities 
$f_1,\ldots,f_d$ can be constructed by
\begin{eqnarray}\label{lee}
f(x_1,\ldots,x_d)=\left\{\prod_{i=1}^df(x_i)\right\}\left\{1+R_{\phi_1,\ldots,\phi_d,\Omega_d}(x_1,\ldots,x_d)\right\},\quad (x_1,\ldots,x_d)\in\mathcal S_1\times\cdots\times \mathcal S_d,
\end{eqnarray}
where $\int_{\mathcal S_i}f_i(x)\phi_i(x)d\mu_i(x)=0$ $\forall i=1,\ldots,d$, 
\begin{eqnarray*}
R_{\phi_1,\ldots,\phi_d,\Omega_d}(x_1,\ldots,x_d)=\sum_{j_1=1}^{d-1}\sum_{j_2=j_1+1}^{d}\delta_{j_1\, j_2}\phi_{j_1}(x_{j_1})\phi_{j_2}(x_{j_2})+\sum_{j_1=1}^{d-2}\sum_{j_2=j_1+1}^{d-1}\sum_{j_3=j_2+1}^{d}\delta_{j_1\, j_2\, j_3}\times \\
\phi_{j_1}(x_{j_1})\phi_{j_2}(x_{j_2})\phi_{j_3}(x_{j_3})+\ldots+\delta_{1\,2\ldots\, d}\prod_{i=1}^d \phi_i(x_i),
\end{eqnarray*}
and $\Omega_d=\left\{\{\delta_{j_1\, j_2}\}_{1\leq j_1<j_2\leq d}, \{\delta_{j_1\, j_2\, j_3}\}_{1\leq j_1<j_2<j_3\leq d},\ldots, \delta_{1\,2\ldots\, d}\right\}$ is the set of parameters controlling the model dependency, which needs to satisfy $1+R_{\phi_1,\ldots,\phi_d,\Omega_d}(x_1,\ldots,x_d)\geq0$ $\forall (x_1,\ldots,x_d)\in\mathcal S_1\times\cdots\times \mathcal S_d$. Note that the proposal by \cite{lee1996} has many parameters to be estimated and which further need to satisfy complicated restrictions to ensure that (\ref{lee}) is a proper joint density function. These restrictions are cumbersome even under low dimensions. For instance, see \cite{bolver2019}, where the restrictions are discussed for a three-dimensional negative binomial distribution based on the extended Sarmanov method, which are tricky to handle when performing inference. For our inferential purposes in this paper, it is crucial to obtain such restrictions in an explicit and simple way.

This motivates us to propose a generalization of the Sarmanov method, which is inspired by the method in \cite{lee1996}. Our proposal aims at parsimonious and explicit and simpler restrictions over the parameters controlling the model dependency ensuring a proper joint density function. Let $f_i$, $\mu_i$, and $\mathcal S_i$ as before for $i=1,\ldots,d$. Then, we propose a $d$-multivariate distribution with marginal densities $f_1,\ldots,f_d$ with respect to the product measure $\mu_1\times\ldots\times\mu_d$ through the joint density function
\begin{eqnarray}\label{multdist}
f(x_1,\ldots,x_d)&=&\left\{\prod_{i=1}^d f_i(x_i)\right\}\left\{\binom{d}{2}+\sum_{j=1}^ {d-1}\sum_{k=j+1}^d\delta_{jk}\phi_j(x_j)\phi_k(x_k)\right\}\bigg/\binom{d}{2}\nonumber\\
&=&\left\{\prod_{i=1}^d f_i(x_i)\right\}\left\{1+\binom{d}{2}^{-1}\sum_{j=1}^ {d-1}\sum_{k=j+1}^d\delta_{jk}\phi_j(x_j)\phi_k(x_k)\right\},
\end{eqnarray}
for $x_1,\ldots,x_d\in\mathcal S_1\times\ldots\times\mathcal S_d$. We have that
\begin{eqnarray}\label{num_aux}
\binom{d}{2}+\sum_{j=1}^{d-1}\sum_{k=j+1}^d\delta_{jk}\phi_j(x_j)\phi_k(x_k)=
\sum_{j=1}^{d-1}\sum_{k=j+1}^d\left\{1+\delta_{jk}\phi_j(x_j)\phi_k(x_k)\right\},
\end{eqnarray}
and therefore the non-negativeness of (\ref{multdist}) can be ensured if all terms of the double summation to the right side of (\ref{num_aux}) are non-negative. This relies on well-known conditions for the bivariate Sarmanov case. For example, in the exponential kernel case, we obtain that (\ref{multdist}) is a proper joint density function if
\begin{eqnarray}\label{mult_range}
\delta_{jk}\in\left(-\dfrac{1}{\max\{(1-\Psi_j)(1-\Psi_k),\Psi_j\Psi_k\}},\dfrac{1}{\max\{\Psi_j(1-\Psi_k),\Psi_k(1-\Psi_j)\}}\right),
\end{eqnarray}
for all $j=1\ldots,d-1$ and $k=j+1,\ldots,d$, where $\Psi_l$ is the Laplace transform at point $\omega$ associated to the marginal density function $f_l$, for $l=1,\ldots,d$. 

\begin{remark}
To conduct Bayesian inference for our model, it is extremely important to know the precise range of the $\delta$'s since we need to check if the draws satisfy the required constraints. Otherwise, such a check could demand a high computational cost. In this paper, we focus on the exponential kernel case, and then the restrictions are given by (\ref{mult_range}). Another important point to reduce the computational cost in what follows is to rewrite the double summation in (\ref{multdist}) in a matrix form. Define $\Phi({\bf x})=(\phi_1(x_1),\ldots,\phi_d(x_d))^\top$, with $\textbf{x}=(x_1,\ldots,x_d)^\top$, and $\boldsymbol\Delta$ being a $d\times d$ matrix with $(j,k)$-th entry given by $\delta_{jk}$ for $j=1\ldots,d-1$ and $k=j+1,\ldots,d$, and the other entries equal to 0. Then, (\ref{multdist}) can be rewritten as
\begin{eqnarray}\label{multdist_matrix}
f(x_1,\ldots,x_d)=\left\{\prod_{i=1}^d f_i(x_i)\right\}\left\{1+\binom{d}{2}^{-1}\Phi({\bf x})^\top{\boldsymbol\Delta}\Phi({\bf x})\right\},\quad x_1,\ldots,x_d\in\mathcal S_1\times\ldots\times\mathcal S_d.
\end{eqnarray}
\end{remark}

\begin{remark}
Our proposed approach assumes that $\delta_{j_1\ldots j_l}=0$ for $l\geq 3$ in (\ref{lee}) to reduce the number of parameters. Another motivation to get these $\delta$'s equal to 0 is to reduce the restrictions on their range. Such a restriction might allow for a limited range of correlation to be captured. The constant $\binom{d}{2}$ considered in (\ref{multdist}) is chosen to get (\ref{num_aux}) so that the restrictions ensuring a proper density function in the Sarmanov bivariate case also guarantee a valid density in the multivariate settup. 
\end{remark}

In the next proposition, we present some quantities of interest related to the joint density function (\ref{multdist}) such as the marginal and conditional density functions. The proof is straightforward and therefore it is omitted.
\begin{proposition}\label{marg_and_cond}
Let $\textbf{X}=(X_1,\ldots,X_d)^\top$ be a random vector with joint probability function (\ref{multdist}). Define $B=\{b_1,\ldots,b_l\}\subset\{1,\ldots,d\}$, $l<d$, $b_1<\ldots<b_l$,  and $C=\{1,\ldots,d\}\setminus B\equiv\{c_1,\ldots,c_{d-l}\}$, $c_1<\ldots<c_{d-l}$. Then, \\
(a) the joint density function of $(X_{b_1},\ldots,X_{b_l})^\top$ assumes the form
\begin{eqnarray*}
f(x_{b_1},\ldots,x_{b_l})=\left\{\prod_{i=1}^l f_{b_i}(x_{b_i})\right\}\left\{1+\binom{d}{2}^{-1}\sum_{j=1}^ {l-1}\sum_{k=j+1}^l\delta_{b_jb_k}\phi_{b_j}(x_{b_j})\phi_{b_k}(x_{b_k})\right\},
\end{eqnarray*}
for $x_{b_1},\ldots,x_{b_l}\in\mathcal S_{b_1}\times\ldots\times\mathcal S_{b_l}$;\\
(b) the conditional density function of $(X_{b_1},\ldots,X_{b_l})^\top$ given $(X_{c_1},\ldots,X_{c_{d-l}})^\top=(x_{c_1},\ldots,x_{c_{d-l}})^\top\in\mathcal S_{c_1}\times\ldots\times\mathcal S_{c_{d-l}}$ is
\begin{eqnarray*}
f(x_{b_1},\ldots,x_{b_l}|x_{c_1},\ldots,x_{c_{d-l}})=\left\{\prod_{i=1}^l f_{b_i}(x_{b_i})\right\}\left\{\dfrac{1+\binom{d}{2}^{-1}\sum_{j=1}^ {l-1}\sum_{k=j+1}^l\delta_{jk}\phi_{j}(x_{j})\phi_{k}(x_{k})}{1+\binom{d}{2}^{-1}\sum_{j=1}^ {d-l-1}\sum_{k=j+1}^{d-l}\delta_{c_jc_k}\phi_{c_j}(x_{c_j})\phi_{c_k}(x_{c_k})}\right\},
\end{eqnarray*}
for $x_{b_1},\ldots,x_{b_l}\in\mathcal S_{b_1}\times\ldots\times\mathcal S_{b_l}$.
\end{proposition}

In the next section, we use our modified Sarmanov method to propose a multivariate count distribution with COM-Poisson marginals.

\section{Multivariate COM-Poisson distribution}\label{mcomp}

Let $X$ be a random variable following a $\mbox{COM-Poisson}(\lambda,\nu)$ distribution with probability mass function defined in (\ref{unicp_pmf}). Then, its moment generating function is given by $E\left(e^{s X}\right) = \dfrac{Z(e^{s}\lambda, \nu)}{Z(\lambda, \nu)}$, for $s\in\mathbb R$, where $Z(\lambda, \nu)$ is defined in (\ref{norm_const}). The mean and variance of $X$ are, respectively, $E(X)=\partial \log Z(\lambda,\nu)/\partial\lambda\equiv \zeta(\lambda,\nu)$ and $\mbox{Var}(X)=\partial \zeta(\lambda,\nu)/\partial\lambda\equiv \zeta'(\lambda,\nu)$. We now use the generalized Sarmanov method discussed in (\ref{multdist}) to propose a $d$-dimensional count distribution having $\mbox{COM-Poisson}(\lambda_i,\nu_i)$ ($\lambda_i,\nu_i>0$; $i=1,\ldots, d$) marginals, which we call the multivariate COM-Poisson (MultCOMP) distribution. For this, we consider the exponential kernel functions $\phi_i(x)=e^{-\omega x}-\dfrac{Z(e^{-\omega}\lambda_i, \nu_i)}{Z(\lambda_i, \nu_i)}$, for $x\in\mathbb N_0$, $\omega>0$, and $i=1,\ldots, d$. 

\begin{definition}
We say that a random vector $\textbf{X}=(X_1,\ldots,X_d)^\top$ follows a multivariate COM-Poisson (MultCOMP) distribution if its joint probability function $P(X_1=x_1,\ldots,X_n=x_n)\equiv f(x_1,\ldots,x_d|\boldsymbol\theta)$ assumes the form
\begin{eqnarray}\label{multcomp}
f(x_1,\ldots,x_d|\boldsymbol\theta)&=&\left\{\prod_{i=1}^d \frac{\lambda_i^{x_i}}{(x_i!)^{\nu_i} Z(\lambda_i,\nu_i)}\right\}\Bigg\{1+\binom{d}{2}^{-1}\sum_{j=1}^ {d-1}\sum_{k=j+1}^d\delta_{jk}\times\nonumber\\
&&\left(e^{-\omega x_j}-\dfrac{Z(e^{-\omega}\lambda_j, \nu_j)}{Z(\lambda_j, \nu_j)}\right)\left(e^{-\omega x_k}-\dfrac{Z(e^{-\omega}\lambda_k, \nu_k)}{Z(\lambda_k, \nu_k)}\right)\Bigg\},\quad x_1,\ldots,x_d\in\mathbb N_0,
\end{eqnarray}
where $\boldsymbol\theta=(\boldsymbol\lambda,\boldsymbol\nu,\boldsymbol\delta,\omega)^\top$, $\boldsymbol\delta=\mbox{vec}\{\delta_{jk}: j=1\ldots,d-1,\, k=j+1,\ldots,d\}$ satisfy (\ref{mult_range}) with $\Psi_i=\dfrac{Z(e^{-\omega}\lambda_i, \nu_i)}{Z(\lambda_i, \nu_i)}$, $i=1,\ldots,d$, $\omega>0$, $\boldsymbol\lambda=(\lambda_1,\ldots,\lambda_d)^\top\in\mathbb R_+^d$, and $\boldsymbol\nu=(\nu_1,\ldots,\nu_d)^\top\in\mathbb R_+^d$. 
\end{definition}

From now on, assume that $\textbf{X}=(X_1,\ldots,X_d)^\top$ is a random vector with joint probability function (\ref{multcomp}), which is denoted by $\textbf{X}\sim\mbox{MultCOMP}(\boldsymbol\lambda,\boldsymbol\nu,\boldsymbol\delta,\omega)$.

For $j\neq k$, we derive the correlation between $X_j$ and $X_k$ to be
\begin{eqnarray}\label{correlation_MultCOMP}
\mbox{corr}(X_j,X_k)=\delta_{jk}\dfrac{\binom{d}{2}^{-1}A_{jk}}{\sqrt{\zeta'(\lambda_j,\nu_j)\zeta'(\lambda_k,\nu_k)}},
\end{eqnarray}
where
\begin{eqnarray*}
A_{jk}\equiv\bigg\{e^{-2\omega}\lambda_j\lambda_k Z'(e^{-\omega}\lambda_j,\nu_j)Z'(e^{-\omega}\lambda_k,\nu_k)+
e^{-\omega}\lambda_j\zeta(\lambda_k,\nu_k)Z'(e^{-\omega}\lambda_j,\nu_j)Z(e^{-\omega}\lambda_k,\nu_k)+e^{-\omega}\lambda_k\zeta(\lambda_j,\nu_j)\\\times   Z'(e^{-\omega}\lambda_k,\nu_k)Z(e^{-\omega}\lambda_j,\nu_j)+\zeta(\lambda_j,\nu_j)\zeta(\lambda_k,\nu_k)Z(e^{-\omega}\lambda_j,\nu_j)Z(e^{-\omega}\lambda_k,\nu_k)\bigg\}\bigg/
\bigg\{Z(\lambda_j,\nu_j)Z(\lambda_k,\nu_k)\bigg\},
\end{eqnarray*}
with $Z'(\lambda,\nu)\equiv \partial Z(\lambda,\nu)/\partial\lambda=\displaystyle\sum_{x=1}^\infty x\lambda^{x-1}/(x!)^\nu$. Since $A_{jk}>0$, the sign of the parameter $\delta_{jk}$ determines if $X_j$ and $X_k$ are negative or positive correlated for $\delta_{jk}<0$ and $\delta_{jk}>0$, respectively. We have that $X_j$ and $X_k$ are independent if $\delta_{jk}=0$. 

\begin{figure}
    \centering
    \includegraphics[width = 0.7\linewidth]{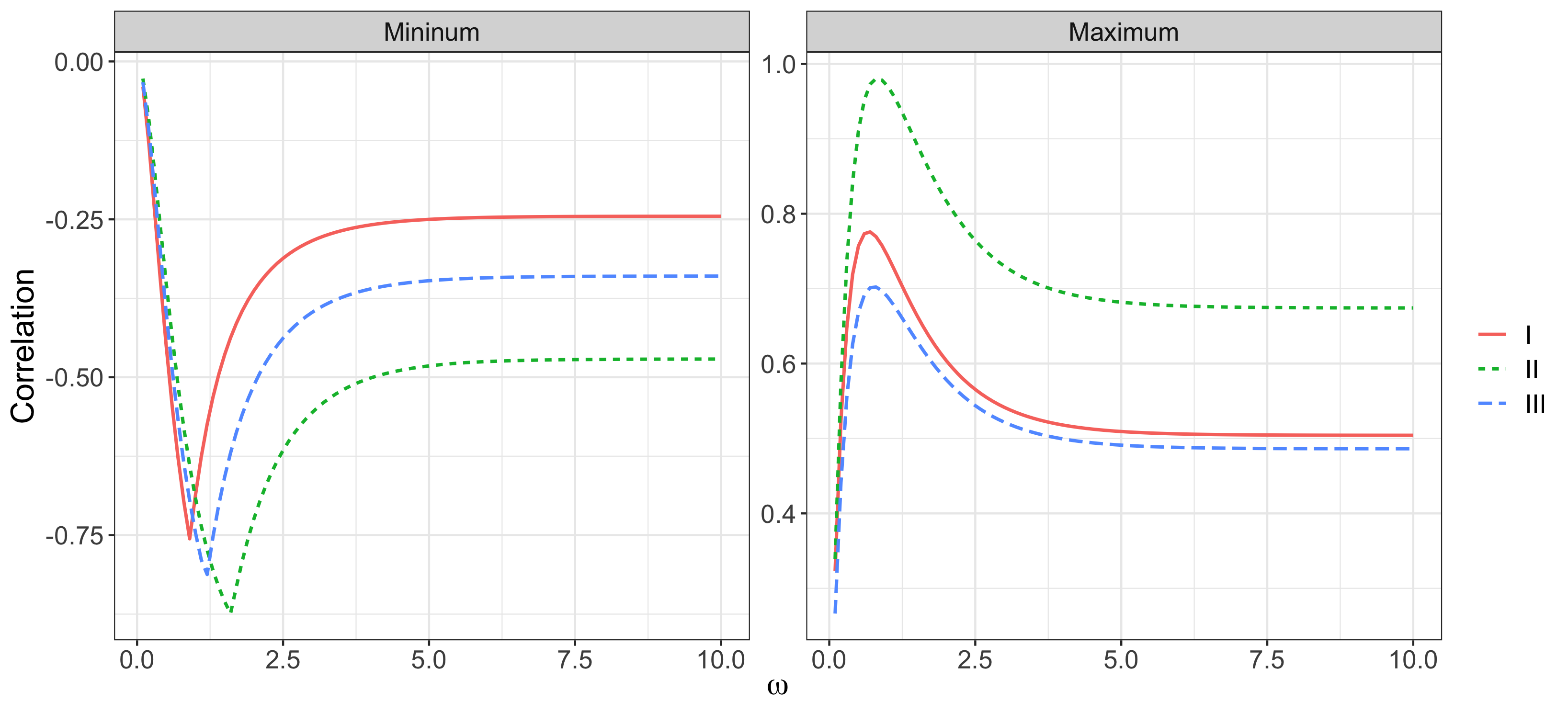}
    \caption{Maximum and minimum correlation as a function of $\omega$ supported by the bivariate COM-Poisson model with $\boldsymbol{\lambda} = (1,1)$ and some values of $\boldsymbol{\nu}$. Solid, dashed, and dotted lines correspond to $\boldsymbol\nu = (0.7, 0.7), (1.5, 1.5)$ and $(0.7,1.5)$, respectively.}
    \label{max_min_cor}
\end{figure}

\begin{remark} An illustration of the correlation supported under the bivariate case is provided in Figure \ref{max_min_cor} which is obtained as follows. For fixed configurations of $\boldsymbol\lambda$ and $\boldsymbol\nu$ we vary $\omega$ and calculate the possible $\delta$ range according to (\ref{mult_range}). Given $\boldsymbol\lambda$, $\boldsymbol\nu$ and $\omega$, there is a linear relationship between $\delta_{jk}$ and the dependency among components $j$ and $k$. Hence, the lower and upper $\delta$ values yield the minimum and maximum correlation under the set configuration. Naturally, the calculation of (\ref{rangecorbiv}) depends on intractable terms under COM-Poisson marginals. In this initial illustration, we replace the infinite summations in the model for truncated ones. More specifically, this is calculated using $\overline{\Psi}_i \equiv \overline{Z}(e^{-\omega}\lambda, \nu)/\overline{Z}(\lambda, \nu)$ for $i=1,2$, where $\overline{Z}(\lambda, \nu)$ denotes replacing \ref{norm_const} with a finite summation in $x=0,\cdots, T$. A minimum $T=1000$ is set, followed by and an iterative procedure that increases $T$ until the difference in successive terms is less than $10^{-5}$. However, in Section~\ref{intractable_terms} we will describe Monte Carlo approaches to handle this intractability. In Figure \ref{max_min_cor}, $\boldsymbol\lambda$ is set to $(1,1)$ and the $\boldsymbol{\nu}$ values vary so that we obtain two overdispersed marginals $\mbox{\textbf{I}}: \boldsymbol{\nu} = (0.7, 0.7)$, both overdispersed $\mbox{\textbf{II}}: \boldsymbol{\nu} = (1.5, 1.5)$ and one of each $\mbox{\textbf{III}}: \boldsymbol{\nu} = (0.7, 1.5)$.
\end{remark}

\begin{remark} This preliminary investigation allows us to conclude the following. The parameter $\omega$ plays the role of increasing the correlation range (with respect to the case $\omega=1$ commonly assumed in the literature) supported by the given $\boldsymbol{\lambda}, \boldsymbol{\nu}$ and $\delta$, but this effect is non-linear. Further, there is evidence that the values associated with the minimum and maximum possible correlation, denoted $\omega_{min}$ and $\omega_{max}$, are not equivalent and vary with the model configuration. For instance, $\omega_{min}=0.9$ and $\omega_{max} =0.7$ in \textbf{I}, while $(\omega_{min}, \omega_{max})$ are (1.6, 0.8) and (1.2, 0.8) under \textbf{II} and \textbf{III} respectively.
\end{remark}

\noindent {\bf Summary of the proposed model features.}  The MultCOMP distribution is a flexible model for analyzing multivariate count data that are dependent since (i) it is defined for a arbitrary dimension $d\in\mathbb N$; (ii) it has a flexible covariance matrix allowing both negative and positive correlations, being the independent case included not at the boundary of the parameter space; (iii) it permits to deal with different degrees of overdispersion and underdispersion for the marginals; (iv) it also allows components to be Poisson, geometric or Bernoulli distributed, with the last two being limiting cases as happens in the univariate COM-Poisson distribution.  

Marginal and conditional probability functions for the MultCOMP law are directly available from Proposition \ref{marg_and_cond}. In what follows, we focus our attention on how to conduct Bayesian inference for the proposed multivariate count model.

\section{Intractable normalising constants and random variable generation}\label{intractable_terms}

Inference and random variable generation for the proposed model depend on being able to evaluate the likelihood (\ref{multcomp}) pointwise. Our likelihood model involves two types of intractable terms, the univariate COM-Poisson normalising constant $\dfrac{1}{Z(\lambda_j, \nu_j)}$ and the ratio $\dfrac{Z(e^{-\omega}\lambda_j,\nu_j)}{Z(\lambda_j, \nu_j)}$ for $j=1,\cdots, d$. We shall denote these respectively by $z^{-1}_j$ and $r_j$, where the dependency on the parameters is suppressed for simplification of notation. This section addresses the estimation of the latter while the former is handled in Section \ref{inference}.

Different methodologies to estimate ratios of normalizing constants of two probability distributions have been developed targeting problems in Bayesian statistics and statistical physics. Quantities of interest are, for example, the Bayes factor and the free energy difference of physical systems. A natural Monte Carlo method for doing this is via Importance Sampling (IS), where draws from the distribution associated with the ratio denominator are taken. The performance of the simple importance sampling scheme will depend on how close the two distributions are. The acceptance ratio, bridge sampling and thermodynamic integration (or path sampling) methods originating from physics are introduced in the statistical literature by \cite{Meng1996} and \cite{gelman1998}. The authors showcase how these methods are linked to the widely known importance sampling, evidencing how they are natural generalizations of it. Other developments are the ratio importance sampling by \cite{torrie1977} (or umbrella sampling) and the annealed importance sampling by \cite{Neal2001}, among others. For a careful assessment of Monte Carlo methods for ratios of normalising constants, we refer the reader to \cite{Chen2000}. In this section, importance sampling and thermodynamic integration estimators of the $r_j$'s are presented and compared via simulation. With these at hand, an algorithm to simulate random draws from the proposed model will be developed.



\subsection{Importance Sampling}\label{IS_estimation}

A simple unbiased estimator of $r\equiv \dfrac{Z(e^{-\omega}\lambda,\nu)}{Z(\lambda, \nu)}$ can be obtained by sampling from a $\mbox{COM-Poisson}(\lambda, \nu)$ distribution. Let $x_1, ..., x_N$ be $N$ independent draws from a $\mbox{COM-Poisson}(\lambda, \nu)$ law. These can be efficiently obtained via the fast-rejection sampler proposed by \cite{benfri2020}. An Importance Sampling (IS) estimator of $r$ is given by
\begin{equation}\label{IS_estimator}
\widehat{r}_{IS} = \frac{1}{N} \sum_{i=1}^{N} \frac{q(x_i|e^{-\omega}\lambda, \nu)}{q(x_i|\lambda, \nu)}.
\end{equation}

The estimator in (\ref{IS_estimator}) is unbiased for the ratio of interest as follows:
\begin{equation*}
    E(\widehat{r}_{IS}) = \frac{1}{N}\sum_{i=1}^{N}  \left(   \sum_{x_i=0}^{\infty} \frac{q(x_i|e^{-\omega} \lambda, \nu)}{q(x_i|\lambda, \nu)} p(x_i|\lambda, \nu) \right) = \frac{1}{N Z(\lambda, \nu)} \sum_{i=1}^{N} \left(   \sum_{x_i=0}^{\infty}q(x_i|e^{-\omega} \lambda, \nu) \right) = \frac{Z(e^{-\omega}\lambda,\nu)}{Z(\lambda, \nu)}.
\end{equation*}

\subsection{Thermodynamic Integration}\label{TINT_estimation}

The Thermodynamic INTegration (TINT) approach is a generalization of the Importance Sampling that has demonstrated successful for many statistical problems. For example, it is the basis of the power-posterior approach for computing Bayesian model evidence \citep{friel2008,friel2014}.

Let $q_0(x)$ and $q_1(x)$ be two unnormalised densities with the same support $\chi$ satisfying $p(x) = q(x)z^{-1}$, where $z = \int q(x) \partial x$. Suppose that it is possible to introduce a class of densities in $\chi$ indexed by a continuous parameter $\alpha$ (with support on some closed interval, say $[0,1]$), say $p(x|\alpha) = q(x|\alpha)/z(\alpha)$, that links the two densities, and we are interested in computing $r = \dfrac{z(1)}{z(0)}$. Thermodynamic integration is also known as path sampling because it relies on creating a path between $q_0(x)$ and $q_1(x)$. One option is to take a geometric path $q(x|\alpha) = q_0(x)^{1-\alpha} q_1(x)^{\alpha}$, $\alpha \in [0,1]$. Having defined a path, we employ the basic identity of path sampling 
\begin{equation}\label{path_sampling}
    \frac{\partial}{\partial \alpha} \log z(\alpha) = E_\alpha \left[ \frac{\partial}{\partial \alpha} \log q(x|\alpha) \right], 
\end{equation}
where the expectation is taken with respect to $p(x|\alpha)$. Integrating $\alpha$ yields the log-ratio of interest once $\log \left(\dfrac{z(1)}{z(0)}\right)  = \displaystyle\int_0^1 E\left[ \frac{\partial}{\partial \alpha} \log q(x|\alpha)  \right]  \partial \alpha$. The continuous parameter $\alpha$ is often called the inverse temperature and is defined such that the path gives us the unnormalised densities of interest at both extremes with the log-ratio resulting from the defined integral. Different strategies to perform thermodynamic integration rely on (i) the definition of the path and (ii) how to perform integration. Common choices for (i) are the geometric and harmonic paths. Regarding (ii), $\alpha$ can be seen as a random variable with a prior distribution, or numerical integration strategies can be adopted.

A thermodynamic integration estimator for the multivariate COM-Poisson intractable ratio is defined by introducing a probability function that is indexed by $\tau$, $p(x|\tau) = q(x|\tau)/z(\tau)$, where $q(x|\tau) = \dfrac{e^{-\tau} \lambda^x}{(x!)^\nu}$, $z(\tau) = \displaystyle\sum_{x=0}^{\infty} q(x|\tau)$ and $\tau \in [0, \omega]$. A geometric path that connects $q(x|\tau = 0) \equiv q_0(x)$ and $q(x|\tau = \omega) \equiv q_\omega(x)$ is $q(x|\tau) = q_0(x)^{-\frac{1}{\omega}(\tau - \omega)} q_\omega(x)^{1 + \frac{1}{\omega}(\tau-\omega)}$ and integration over the inverse temperature $\tau$ yields the desired log-ratio $\dfrac{z(\omega)}{z(0)}$. Following the notation in \cite{gelman1998}, we denote $U(x, \tau)\equiv\log q(x|\tau)=\omega^{-1}(\log q_\omega(x) - \log q_0(x))$. The path sampling identity gives us that $\log r = \displaystyle\int_0^{\omega} E_\tau \left[ U(x, \tau) \right] \partial \tau$, where the expectation is taken with respect to a COM-Poisson($\lambda e^{-\tau}, \nu$) distribution. If we define $\tau$ to be a random variable with density $p(\tau)$, a Monte Carlo estimator of $\log r$ is $\widehat{\log r} = \dfrac{1}{N} \displaystyle\sum_{i=1}^N \dfrac{U(x, \tau_i)}{p(\tau_i)}$. For example, we can consider $\tau \sim U[0, \omega]$ and sample from $p(x, \tau) = p(x|\tau)p(\tau)$. 

A thermodynamic estimator that sets the inverse temperature to be a random variable can result in a poor performance if values on the extremes of the interval are not sampled frequently enough under $p(\tau)$ \citep{friel2014}. A popular alternative is to adopt numerical integration over a discretised range $0 = \tau_1 < \tau_2< \ldots < \tau_{n_{rungs}} = \omega$, guaranteeing proper exploration of $\tau$ values. An estimator based on the trapezoid rule is given by 
\begin{eqnarray*}
\widehat{\log r} = \sum_{i=1}^{n_{rungs}} (\tau_{i} - \tau_{i-1}) \frac{E_{\tau_{i-1}}\left[ U(x, \tau_{i-1}) \right]   +  E_{\tau_i}\left[ U(x, \tau_i) \right] }{2}.
\end{eqnarray*}

At each $\tau_i$, a number of independent draws from a $\mbox{COM-Poisson}(e^{-\tau_i} \lambda, \nu)$ distribution are used to estimate expectations as Monte Carlo averages.  A TINT estimator employing numerical integration relies additionally on (i) the number and schedule of the discretisation terms (also called rungs) and (ii) the number of simulated draws per rung. Finding an optimal form for (i) is a non-trivial problem for which the recommendation is to adopt a power fraction schedule \citep{oates2014}. Under this approach, $\tau$ values are placed according to $(i/n_{rungs})^c$ with $c>1$ and $i =1,\ldots, n_{rungs}$. Schedules of this form have commonly been adopted in literature, demonstrating to be a successful choice \citep{friel2008}. We adopt $\tau_i = \omega (i/n_{rungs})^5$ and $n_{rungs}$ depending on the interval length $\omega$. Simulation studies are performed in the next subsection to compare the performance of the IS, TINT with prior distribution (TINT-prior) and numerical integration TINT (TINT-trapezium) estimators for $r$.

\subsection{Comparing ratio estimators}\label{r_comparison}

Estimators for $r$ based on the IS and TINT methods are compared in this section via simulation. For this task, we consider the pairs $(\lambda,\nu)=(1.5,1), (1,0.5)$ and $\omega = 0.5, 3$. For a fair comparison, approximately the same total number of draws ($n_{total}$) is used when computing $\widehat{r}$ through the alternative estimators. Effectively, this means that IS and TINT-prior use $n_{total}$ draws and $n_{total}$ is spread over the grid of $\tau$ values under TINT-trapezium. Two discretisations are assessed for the latter. The first takes $n_{rungs} = \lceil \omega/0.1 \rceil$ and the second $n_{rungs} = \lceil \omega/0.05 \rceil$
with the number of draws per rung being $\lceil n_{total}/n_{rungs} \rceil$. We explore a grid of $n_{total}$ from 10K to 200K draws where 200 replications are used to compute the Monte Carlo standard deviation of each estimator.

\begin{figure}
    \centering
    \includegraphics[width=0.8\linewidth]{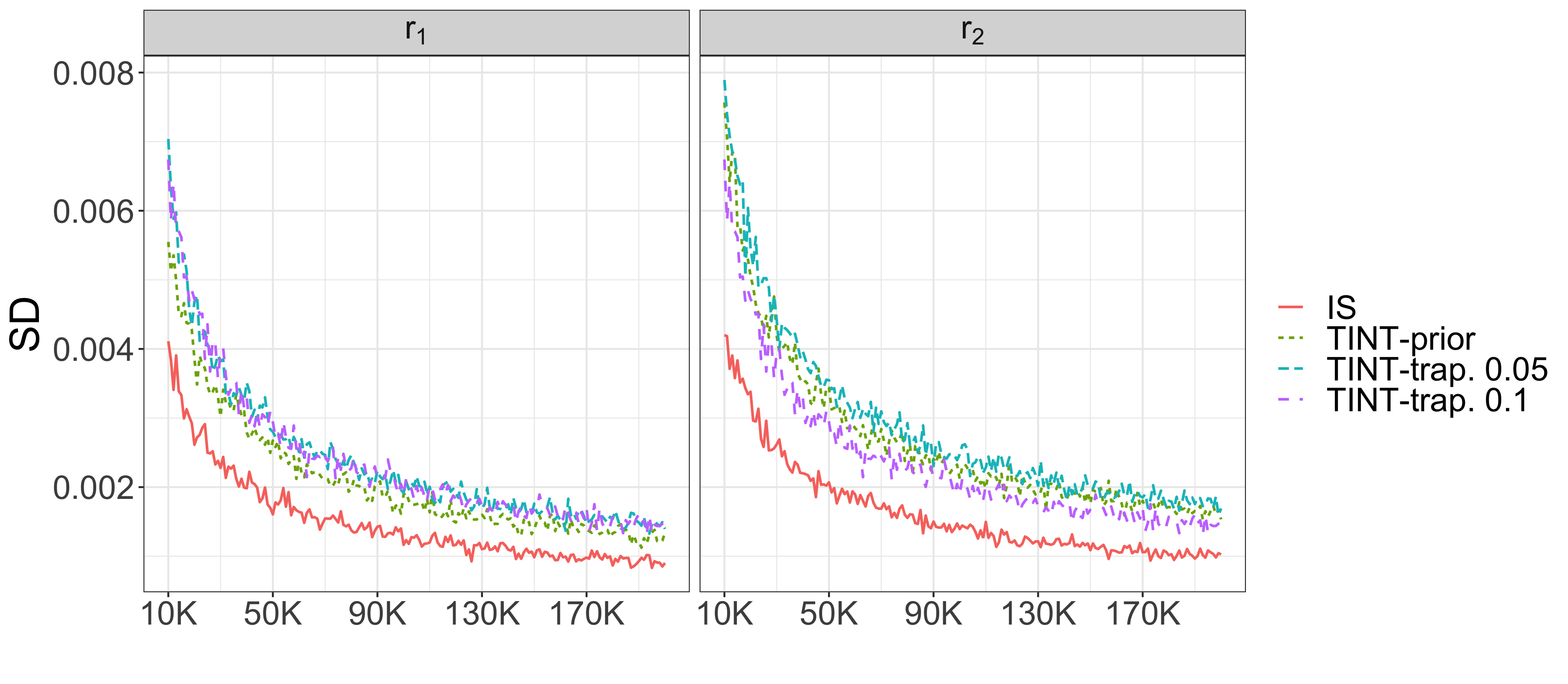}
    \caption{Monte Carlo standard deviation versus total number of draws of $r_1$ and $r_2$ estimated with importance sampling (IS) and variations of thermodynamic integration (TINT). Under TINT-prior, a uniform prior is assumed for the inverse temperature, and TINT-trapezium employs numerical integration using the trapezoid rule with 0.1 or 0.05 spacing. }\label{plot_sd_methods_omega3}
\end{figure}

Results for $\omega =3$ with $r_1 = \dfrac{Z(e^{-3} 1.5, 1)}{Z(1.5, 1)}$ and $r_2 = \dfrac{Z(e^{-3} , 0.5)}{Z(1, 0.5)}$ are reported in Figure \ref{plot_sd_methods_omega3} and those due to $\omega =0.5$ can be found in the Supplementary Material, which can be obtained from the authors upon request. In both experiments, it is shown that the smallest variability is due to the simplest estimation procedure (IS) over the entire $n_{total}$ range. This indicates that the importance density $\mbox{COM-Poisson}(e^{-\omega} \lambda, \nu)$ is close enough to $\mbox{COM-Poisson}(\lambda, \nu)$ to yield a low variability even for high $\omega$. 

\subsection{Multivariate COM-Poisson sampler}

It is now possible to introduce an algorithm to sample from the proposed distribution. For simplicity, denote $r_j = \dfrac{Z(e^{-\omega}\lambda_j, \nu_j)}{Z(\lambda_j, \nu_j)}$. Our strategy is to draw $X_1$ from its marginal and sample from the sequence of conditional distributions $p(x_j|x_{j-1},\cdots, x_1)$, for $j = 2, \cdots, d$. By defining $\widetilde{p}(x_j|x_{j-1},\cdots, x_1) = Z(\lambda_j, \nu_j) p(x_j|x_{j-1},\cdots, x_1)$, we obtain an unnormalised probability function that depends on $\boldsymbol{r}\equiv (r_1,\ldots,r_d)^\top$ but no longer on $\boldsymbol{z}^{-1}\equiv(z_1^{-1},\ldots,z_d^{-1})^\top$. An estimator for $\widetilde{p}(x_j|x_{j-1},\cdots, x_1)$ is obtained by plugging in $\boldsymbol{\widehat{r}}$. If we evaluate $\widetilde{p}(x_j=k|x_{j-1},\cdots, x_1)$ with $k \in \{0,\cdots,K\}$ for sufficiently large $K$, a normalised probability function can be recovered as $\widetilde{p}(x_j=k|x_{j-1},\cdots, x_1)/\sum_{k=0}^K \widetilde{p}(x_j=k|x_{j-1},\cdots, x_1)$. In other words, $\widehat{Z}(\lambda_j, \nu_j)$ is estimated using the fact that for sufficiently large $K$
 \begin{align}
 \sum_{k=0}^{K} \widetilde{p}(x_j = k|x_{j-1},\cdots, x_1) = Z(\lambda_j, \nu_j)  \underbrace{ \sum_{k=0}^{K} p(x_j =k|x_{j-1},\cdots, x_1)}_{\approx1}. \nonumber
 \end{align}

In Algorithm \ref{bcp_sampler}, we provide a pseudo-code to draw $N$ independent samples following approximately a $\mbox{MultCOMP}(\boldsymbol{\lambda}, \boldsymbol{\nu}, \boldsymbol{\delta}, \omega)$ distribution. The quality of approximation depends on (i) how well we estimate $\boldsymbol{r}$ and (ii) the choice of $K$. From Section \ref{r_comparison}, we recommend the use of the importance sampling estimator with $n_{draws}$ over 130K, region where the standard deviation seems to stabilize. For (ii) we adopt the strategy of setting a minimum value of $K$ that is increased until the difference of successive probabilities is less than a pre-specified tolerance.

\begin{algorithm}
\small
\KwIn{$N, K, \boldsymbol{\lambda}, \boldsymbol{\nu}, \boldsymbol{\delta}, \omega$}

\For{i $\leftarrow 1:N$}{

Sample $X_1 \sim \mbox{COM-Poisson}(\lambda_1, \nu_1)$. 

\For{j $\leftarrow 2:d $}{
Estimate $\widehat{r}_j$.

Calculate $\widetilde{p}(x_j = k|x_{j-1},\cdots, x_1)$ for $k \in \{0,\cdots,K\}$, where $K$ is sufficiently large. 

Sample from $\{0,\ldots,K\}$ with probabilities $\dfrac{\widetilde{p}(x_j = k|x_{j-1},\ldots, x_1)}{\sum_{k=0}^K \widetilde{p}(x_j = k|x_{j-1},\ldots, x_1)}$.

}

}
\KwOut{$N$ draws following approximately a $d$-dimensional multivariate COM-Poisson distribution.}

\caption{Multivariate COM-Poisson sampler}\label{bcp_sampler}
\end{algorithm}

\section{Doubly-intractable Bayesian inference}\label{inference}

First coined by \cite{murray2006}, the term \textit{doubly-intractable} refers to the posterior distribution of a Bayesian model involving an intractable likelihood function. It refers to the fact that, additionally to the model evidence which is already intractable in most problems, there is a normalising constant in the model likelihood that is not analytical and depends on the model parameters. Intractability of the likelihood commonly arises from the presence of latent random variables that are not straightforward to integrate, or it simply might be difficult to calculate the normalising constant. For example, this difficulty can be due to dimensionality as in exponential random graph models, Gibbs random fields and permutation models, just to name a few. The normalizing constant of these models require the evaluation of all possible values of the random variable which is non-trivial except from very small graphs and permutations of a small number of items.

Bayesian inference of intractable likelihood problems require special attention as standard Markov Chain Monte Carlo 
(MCMC) methods such as the Metropolis-Hastings algorithm are not suitable as they depend on point-wise evaluation of the likelihood function. Different approaches that bypass this issue exist with some examples being the usage of composite-likelihoods \citep{varin2011} also known as pseudo-likelihoods, or even likelihood-free methods such as Approximate Bayesian Computation (ABC) \citep{abc_handbook}. We focus on a class of MCMC algorithms that have been proposed for doubly-intractable problems. These can be classified into asymptotically exact and asymptotically inexact (or noisy) algorithms, depending on whether their stationary distribution is the target posterior exactly or approximately \citep{park2018}, \citep{alq:fri:eve:bol14}.

Assume that $\boldsymbol{X}| \boldsymbol\theta$ follows a MultCOMP model with joint probability function assuming the form (\ref{multcomp}) and the parameter vector denoted by $\boldsymbol\theta = (\boldsymbol{\lambda}, \boldsymbol{\nu}, \boldsymbol{\delta},\omega)^\top$, which is assumed to follow a prior distribution. Then, $p(\boldsymbol\theta| \boldsymbol{X}) \propto p(\boldsymbol{X}| \boldsymbol\theta) p(\boldsymbol\theta)$ is the posterior model, which is doubly-intractable. In this section, two MCMC methods based on auxiliary variables are developed to perform inference for the proposed multivariate count model.

\subsection{Pseudo-marginal inference}

Here we consider Pseudo-marginal MCMC, an approach which uses an unbiased estimator $\widehat{p}(\boldsymbol{X}|\boldsymbol\theta)$ of the likelihood function. Pseudo-marginal methods introduce auxiliary variables $\boldsymbol{Y}$ that aim to facilitate the approximation of the intractable posterior which is then used in the Metropolis-Hastings acceptance rate. Introduced in the context of genetics, pseudo-marginal methods require a positive and unbiased likelihood estimator, with two possible implementations studied by \cite{andrieu2009}. These are named "Monte Carlo Within Metropolis" (MCWM) and "Grouped Independence Metropolis Hastings" (GIMH), which differ in terms of how the auxiliary draws are used. While MCWM refreshes $\boldsymbol{Y}$ at the current and proposed parameter values of each iteration, GIMH recycles the draws using $\widehat{p}(\boldsymbol\theta|\boldsymbol{X})$ of when $\boldsymbol\theta$ was last accepted. Convergence properties studied in \cite{andrieu2009} state that the Markov chain resulting from MCWM does not have the desired invariant distribution, while its variant GIMH targets $p(\boldsymbol\theta, \boldsymbol{Y}|\boldsymbol{X})$ as desired. For this reason, the application of pseudo-marginal methods in the literature have commonly focused on the GIMH.

Suppose that a likelihood estimator $\widehat{p}(\boldsymbol{X}|\boldsymbol\theta)$ is available provided ${\bf Y}\equiv(y_1,\ldots, y_N)^\top$ independent draws from an auxiliary density $g_{\boldsymbol\theta}(\cdot)$, where the subscript denotes the dependence of the auxiliary density on $\boldsymbol\theta$. A general framework for conducting GIMH is given in Algorithm \ref{GIMH}.

\begin{algorithm}
\small
\KwIn{$\boldsymbol\theta^{(t)}$, initial draws $\boldsymbol{Y} \sim g_{\boldsymbol\theta^{(t)}}(\cdot)$, $\boldsymbol{X}$}

Propose $\boldsymbol\theta' \sim h(\cdot|\boldsymbol\theta^{(t)})$;

Draw  $\boldsymbol{Y}' \sim g_{\boldsymbol\theta'}(\cdot)$;

Estimate  $\widehat{p}(\boldsymbol{X}|\boldsymbol\theta')p(\boldsymbol\theta')$;

Compute the acceptance probability $\pi = \min \left\{ 1, \frac{h(\boldsymbol\theta^{(t)}|\boldsymbol\theta') \widehat{p}(\boldsymbol{X}|\boldsymbol\theta')\pi(\theta')}{h(\boldsymbol\theta^{(t)}|\boldsymbol\theta') \widehat{p}(\boldsymbol{X}|\boldsymbol\theta^{(t)})p(\boldsymbol\theta^{(t)})} \right\}$;

With probability $\pi$ set
$\boldsymbol\theta^{(t+1)} = \boldsymbol\theta'$ and $\boldsymbol{Y}^{(t+1)} = \boldsymbol{Y}'$, otherwise $\boldsymbol\theta^{(t+1)} = \boldsymbol\theta^{(t)}$ and $\boldsymbol{Y}^{(t+1)} = \boldsymbol{Y}^{(t)}$;

\caption{Grouped Independence Metropolis-Hastings (GIMH)}\label{GIMH}
\end{algorithm}

An unbiased likelihood estimator of the joint probability function given in (\ref{multcomp}) can be obtained by estimating unbiasedly and independently $r_j$ and $z^{-1}_j$, for $j=1, \cdots, d$. We leverage the IS estimator introduced in Subsection \ref{IS_estimation} for the first and the latter can be handled via the method proposed by \cite{benfri2020} in the context of univariate COM-Poisson regression. By sampling $N$ draws of a $\mbox{COM-Poisson}(\lambda, \nu)$ distribution through their fast-rejection sampler, an unbiased estimator of $\dfrac{1}{Z(\lambda, \nu)}$ is given by $\dfrac{\widehat{M}^{(N)}}{B}$, where $\widehat{M}^{(N)}$ is the ratio of $n_N$, the number of draws required for $N$ acceptances, and $N$. The denominator $B$ is the envelope's tractable bound which is the normalising constant of a Poisson  or geometric distributions in the respective cases when $\nu\geq 1$ and $\nu < 1$.

Hence, $\widehat{p}(\boldsymbol{X}|\theta)$ relies on drawing two sets of auxiliary variables from univariate $\mbox{COM-Poisson}(\lambda_j,\nu_j)$ distributions using the fast-rejection sampler for $j=1,\cdots,d$. One set is used to estimate $\widehat{r}_j$ via IS and the other yields $\widehat{z}_j^{-1}$ via the envelope's acceptance. Evidently, the computational cost could be reduced if the same draws are used to compute both $\widehat{z}_j^{-1}$ and $\widehat{r}_j$ but this would introduce dependency among the estimators. Consequently, it would be required to show unbiasedness of $E[\widehat{z}_j^{-1}  \widehat{r}_j]$ which is not straightforward or even not true. By simulating two separate sets, we are able to guarantee an unbiased estimator of $\widehat{p}(\boldsymbol{X}|\boldsymbol\theta)$ since $z^{-1}_j$ and $\widehat{r}_j$ are independent and unbiased. Another advantage in this approach is that different number of draws for $\widehat{z}^{-1}$ and $\widehat{r}$, to be denoted $N_z$ and $N_r$, can be set if a distinct precision is required.

We investigate GIMH mixing and how it is affected by the different quantities being estimated in the next section. This will guide the choices of $N_z$ and $N_r$. First, a GIMH algorithm with single-site updates for the multivariate COM-Poisson posterior model is described given $N_z$ and $N_r$. After specifying initial states for the parameters, we store the $N_z \times d$ and $N_r \times d$ matrices of auxiliary draws that are used to compute $\boldsymbol{\widehat{z}}^{-1}$ and $\boldsymbol{\widehat{r}}$. These are denoted by $\underline{\boldsymbol{Y}}_z$ and $\underline{\boldsymbol{Y}}_r$, respectively. The update of one $\lambda_j$ follows from proposing a new state $\lambda_j'$ and drawing the auxiliary variables at the proposed parameter value to compute $\widehat{r}_j'$. In Algorithm \ref{GIMH_mcp}, the notation $\boldsymbol Y_z^{(t)}[\_,j] \leftarrow \boldsymbol{Y}_{z,j}'$ denotes the update of the $j^{(t)}$ column of the current $\boldsymbol{Y}_{z,j}^{(t)}$ matrix with auxiliary draws taken at the proposed parameter value, $\boldsymbol Y_{z,j}' \sim \mbox{COM-Poisson}(\lambda_j', \nu_j^{(t)})$, denoted by $\boldsymbol{ \underline{ Y}}_{z,j}'$. This is similar for $\underline{ \boldsymbol{Y}}_{r, j}'$. If the proposed stated is accepted, $\underline{\boldsymbol{Y}}_z$ and $\underline{\boldsymbol{Y}}_r$ are updated to $\underline{\boldsymbol{Y}}_z'$ and $\underline{\boldsymbol{Y}}_r'$. Algorithm \ref{GIMH_mcp} details the update of the location parameter vector $\boldsymbol{\lambda}$ with  those of $\boldsymbol{\nu}$ following in a similar manner.

\begin{algorithm}

\small

\KwIn{$N_z$, $N_r$, $\boldsymbol\theta^{(t)}$, $\underline{\boldsymbol{Y}}_r^{(t)}$, $\underline{\boldsymbol{Y}}_z^{(t)}$, $\widehat{\boldsymbol{z}}^{{-1}^{(t)}}$, $\widehat{\boldsymbol{r}}^{(t)}$, $\boldsymbol{X}$}


\For{$j \rightarrow 1:d$}{

    Propose $\lambda_j' \sim \mbox{Log-Normal}(\log\lambda_j^{(t)}, \sigma_{\lambda_j})$ and set $\boldsymbol{\lambda}' = \boldsymbol{\lambda}^{(t)}_{[j \leftarrow \lambda_j']}$;
    
    Set $\boldsymbol\theta' = (\boldsymbol{\lambda}', \boldsymbol{\nu}^{(t)}, \boldsymbol{\delta}^{(t)}, \omega^{(t)})$;
    
    Simulate $N_z$ draws $\boldsymbol{Y}_{z,j}' \sim \mbox{COM-Poisson}(\lambda_j', \nu_j^{(t)})$ and set $\underline{\boldsymbol{Y}}_z' \equiv \boldsymbol{Y}_z^{(t)}[\_, j] \leftarrow \boldsymbol{Y}_{z,j}'$;
    
    Simulate $N_r$ draws $\boldsymbol{Y}_{r,j}' \sim \mbox{COM-Poisson}(\lambda_j', \nu_j^{(t)})$ and set $\underline{\boldsymbol{Y}}_r' \equiv \boldsymbol{Y}_r^{(t)}[\_, j] \leftarrow \boldsymbol{Y}_{r,j}'$;
    
    Use $\boldsymbol{Y}_{z,j}'$ to calculate $\widehat{z'}_j^{{-1}}$ and set $\boldsymbol{\widehat{z'}}_j^{{-1}} \equiv {\boldsymbol{\widehat{z}}_j^{-1}}^{(t)}[j] \leftarrow \widehat{z'}_j^{{-1}} $;
    
    Use $\boldsymbol{Y}_{r,j}'$ to calculate $\widehat{r}_j'$ and set $\boldsymbol{\widehat{r}}_j' \equiv \boldsymbol{\widehat{r}}_j^{(t)}[j ] \leftarrow \widehat{r}_j'$;
    
    Check if $\lambda_j'$ yields a valid pmf by ensuring that every component of $\boldsymbol\delta$ satisfies (\ref{mult_range}). If this is not achieved, return to the Step 2.
    
    Calculate $\pi = \min \left\{1, \frac{p(\lambda_j') \widehat{p}(\boldsymbol{X}|\boldsymbol\theta',\widehat{\boldsymbol{z}'}^{-1}, \widehat{\boldsymbol{r}}') h(\lambda_j^{(t)}|\lambda_j')  }{p(\lambda_j^{(t)}) \widehat{p}(\boldsymbol{X}|\boldsymbol\theta^{(t)},\widehat{\boldsymbol{z}}^{-1}_{(t)}, \widehat{\boldsymbol{r}}^{(t)}) h(\lambda_j'|\lambda_j^{(t)})  } \right\}.$
    
    With probability $\pi$ set $\lambda_j^{(t+1)} = \lambda_j'$, $\widehat{\boldsymbol{z}}^{-1}_{(t+1)} = \widehat{\boldsymbol{z}'}^{-1}$, $\widehat{\boldsymbol{r}}^{(t+1)} = \widehat{\boldsymbol{r}}'$ and $\boldsymbol{\underline{Y}}_z^{(t+1)}  = \boldsymbol{\underline{Y}}_z'$, $\boldsymbol{\underline{Y}}_r^{(t+1)}  = \boldsymbol{\underline{Y}}_r'$.
}

\caption{GIMH for multivariate COM-Poisson model: $\boldsymbol{\lambda}$ update}\label{GIMH_mcp}
\end{algorithm}

The Step 8 in \ref{GIMH_mcp} is implemented to ensure that the current $\boldsymbol\delta$ values are comprised by the interval implied by the proposed parameter value (interval given in (\ref{mult_range})). Updates of $\omega$ and $\boldsymbol{\delta}$ differ on the usage of the auxiliary draws. Since their distribution does not depend on $\omega$ or $\boldsymbol{\delta}$, no new draws are required at these moves. A new $\omega$ value is proposed from the same log-normal kernel and recycled draws are used to recompute all $\boldsymbol{\widehat{r}}$ elements. Finally, each $\delta_{jk}'$ is simulated from a truncated-normal($\delta_{jk}^{(t)}$, $\sigma_{\delta_{jk}}$) distribution in the interval (\ref{mult_range}), for  $j =1,\cdots, d-1, k = j+1,\cdots,d$.

\subsubsection{Mixing aspects of GIMH}\label{mixing_gimh}

This section is dedicated to investigating the effects of $\widehat{\boldsymbol{r}}$ and $\widehat{\boldsymbol{z}}^{-1}$ in the performance of the GIMH. This is crucial as the GIMH chain can get stuck in regions of the parameter space if the likelihood is substantially overestimated at any given iteration \citep{drovandi2018}, resulting in poor mixing. Naturally, the precision to which the likelihood is estimated depends on how well we estimate $\widehat{\boldsymbol{r}}$ and $\widehat{\boldsymbol{z}}$. In this section, we are able to investigate separately how the variability associated to $\widehat{\boldsymbol{r}}$ and $\widehat{\boldsymbol{z}}$ affect the likelihood estimator, and hence the overall GIMH mixing. This is possible because the log-likelihood function of one single $d$-dimensional multivariate COM-Poisson observation $\boldsymbol{X}$ can be decomposed in the summation of two terms. The first is contribution due to the marginal univariate COM-Poisson distributions and the second one is the dependency part introduced by the Sarmanov method with exponential kernel. This is indicated in the next equation where the marginal term depends solely on $\widehat{\boldsymbol{z}}^{-1}$ and the second on $\widehat{\boldsymbol{r}}$:
\begin{eqnarray}
\small
\log f({\bf X}| \boldsymbol{\theta}) =  \underbrace{ \sum_{j=1}^{d} \left\{  x_j \log(\lambda_j) -\nu_j \log(x_j!) + \log (\widehat{z}^{-1}_j ) \right\}}_{\mbox{Marginal}} + 
 \underbrace{ \log \left\{ 1 + {d \choose 2}^{-1} \sum_{l=1}^{d-1} \sum_{k = l+1}^{d} \delta_{lk}\left( e^{-\omega x_l}  -\widehat{r}_l  \right) \left( e^{-\omega x_k}  -\widehat{r}_k  \right)  \right\} }_{\mbox{Kernel}}. \nonumber
\end{eqnarray}

When considering a sample of $n$ observations, the contribution of $\widehat{\boldsymbol{z}}^{-1}$ is $\sum_{j=1}^d n \widehat{z}^{-1}_j$, hence there is a multiplicative effect of the sample size for each $\widehat{z}^{-1}_j$, and the overall influence of this estimator also increases with $d$. We begin by fixing $N_z = N_r = 10K$ running GIMH while storing the marginal and kernel contributions at each iteration. Resulting trace plots give an idea of how the GIMH mixing associated to each type of estimator. Results reported in the section are due to 50K iterations for a simulated data set of $500$ tri-dimensional observations with $\boldsymbol{\lambda} = (1.5, 1, 0.5)$, $\boldsymbol{\nu} = (1, 0.5, 0.8)$, $\omega = 3$, $\delta_{12} = 3.5$, $\delta_{13}= -2.5$, and $\delta_{23} = -3$.

\begin{figure}
\centering
\begin{subfigure}{.55\textwidth}
  \centering
  \includegraphics[width=1\linewidth]{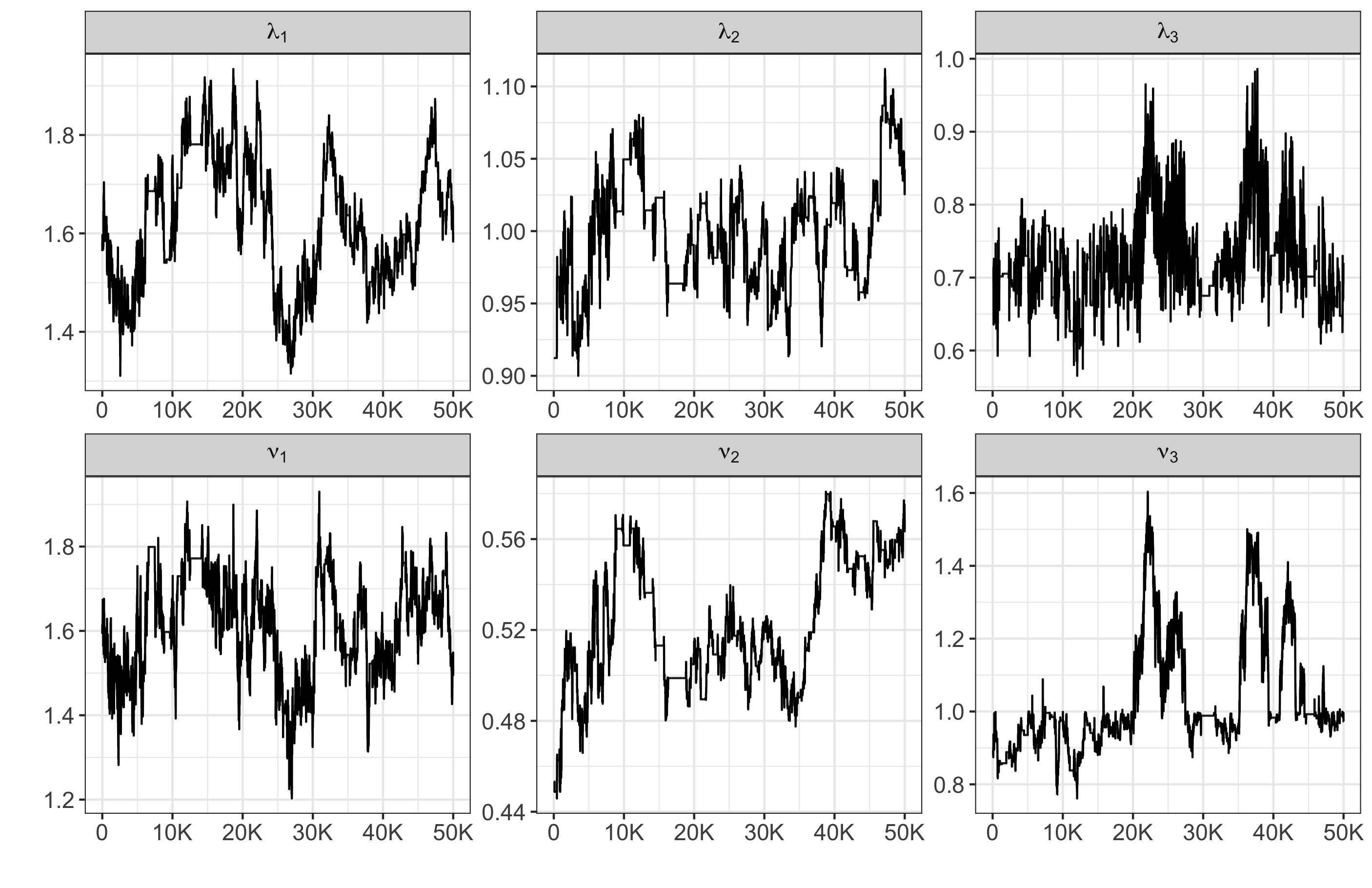}
  \label{fig:sub1}
\end{subfigure}%
\begin{subfigure}{.45\textwidth}
  \centering
  \includegraphics[width=1\linewidth]{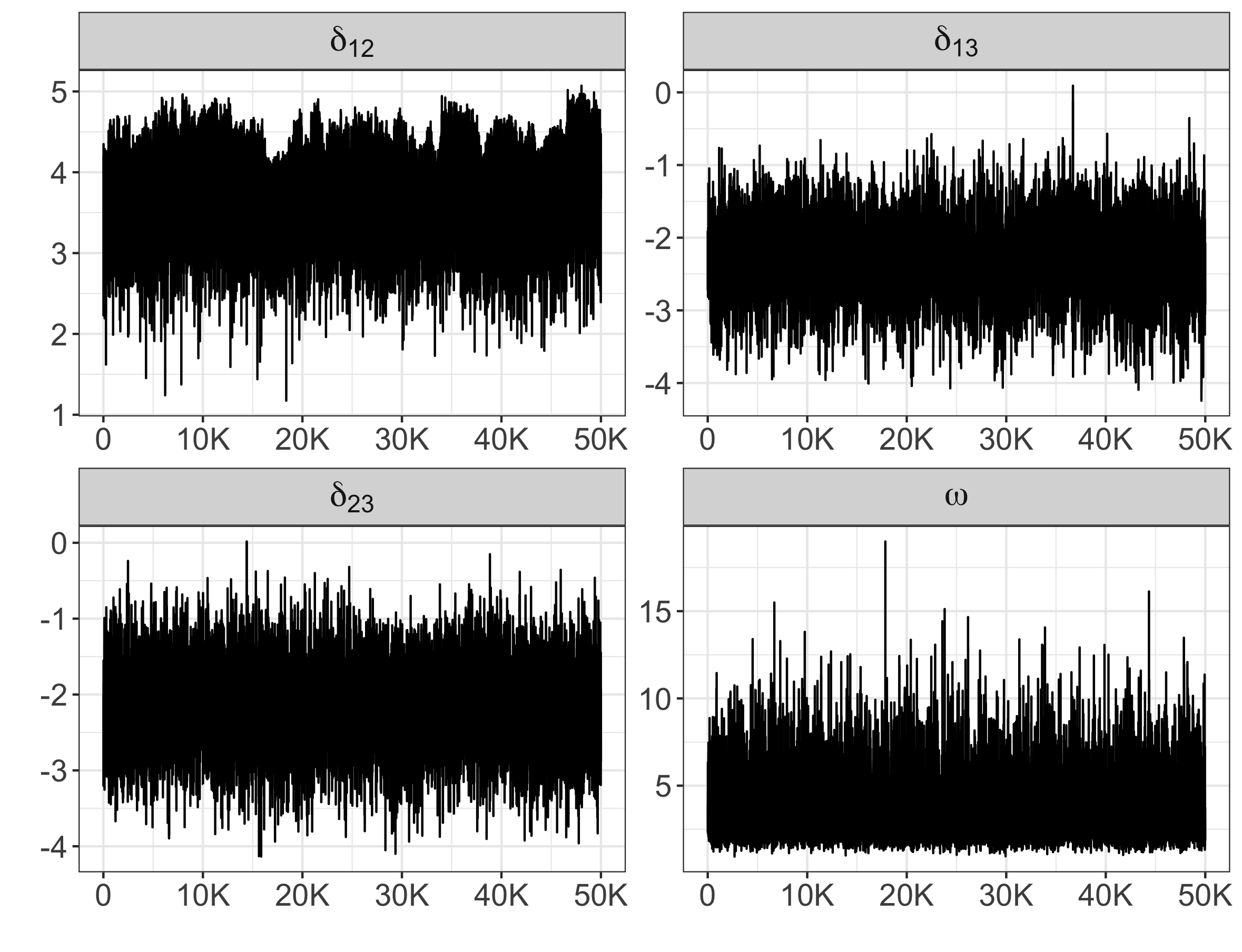}
  \label{fig:sub2}
\end{subfigure}
\caption{Trace plot of the trivariate COM-Poisson model parameters obtained under GIMH approach with $N_z=N_r=10K$ and sample size $n=500$.}
\label{GIMH_convergence_n500}
\end{figure}

\begin{figure}
    \centering
    \includegraphics[width=0.7\linewidth]{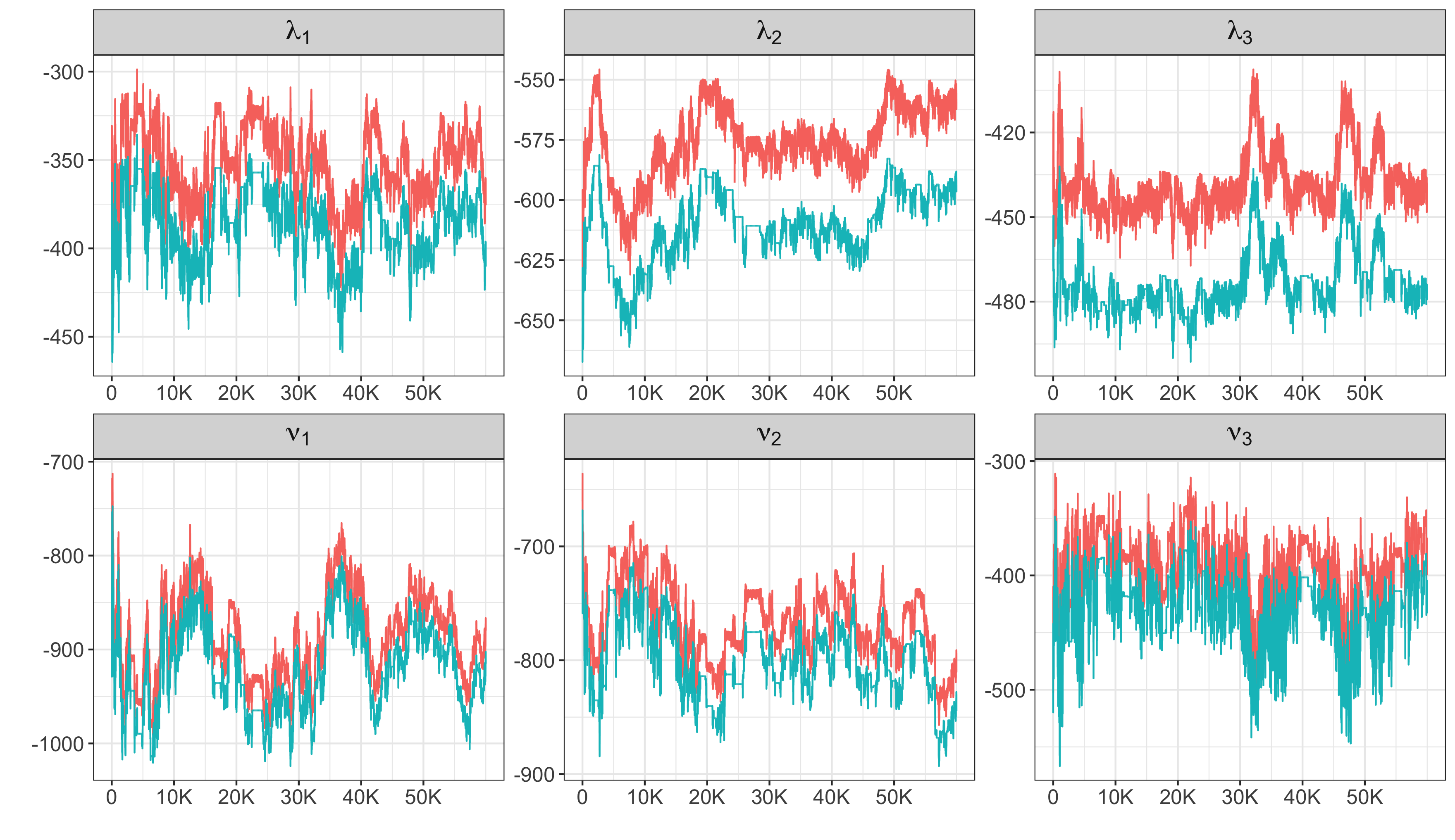}
    \caption{Trace plot of the log-likelihood function up to proportionality (red line) and its marginal contribution (blue line) for the trivariate COM-Poisson parameters that depend on both $\widehat{\boldsymbol{z}}^{-1}$ and $\boldsymbol{\widehat{r}}$.}
    \label{trace_decomposition}
\end{figure}

Figure \ref{GIMH_convergence_n500} displays trace plots of the posterior model parameters evidencing bad mixing for the $\boldsymbol{\lambda}$ and $\boldsymbol{\nu}$ chains while those of $\boldsymbol{\delta}$ and $\omega$ behave well. The log-likelihood function proportional to $\boldsymbol{\delta}$ and $\omega$ is given only by the kernel term, while those of $\boldsymbol{\lambda}$ and $\boldsymbol{\nu}$ depend on both type of estimators. This suggests that $N_r = 10K$ is sufficient for $\widehat{\boldsymbol{r}}$ but a higher precision is required for $\widehat{\boldsymbol{z}}^{-1}$. To confirm the hypothesis that the poor mixing in $\boldsymbol{\lambda}$ and $\boldsymbol{\nu}$ is caused by $\boldsymbol{\widehat{z}}^{-1}$, we plot in Figure \ref{trace_decomposition} the parameters total proportional log-likelihood functions and its contribution due to the marginal part. The kernel term cannot be decomposed in this manner so it is the same for all $\boldsymbol{\lambda}$ and $\boldsymbol{\nu}$.

Figure \ref{trace_decomposition} shows how the largest part of the proportional log-likelihood function is given by the marginal contribution. Consequently, these parameters are largely influenced by $\widehat{\boldsymbol{z}}^{-1}$. According to \cite{doucet2015}, for good performance of the GIMH, the log-likelihood function should be estimated with a standard deviation between 1 and 1.7. More specifically, for the case in which the efficiency of the Metropolis-Hastings algorithm using the exact likelihood is unknown, the suggested value is 1.2. Assumptions in this work are that the noise introduced by the log-likelihood estimator is Gaussian with variance inversely proportional to the number of samples used to construct the estimator and it is also independent of the parameter value. We implement this strategy by setting an adaptation phase for $N_z$ before running the MCMC chain. Naturally, it is also possible to calibrate each $N_{z,j}, j =1,\cdots, d$ but we find that taking $N_{z,j} = N_z\,\, \forall j$ is a simpler and more conservative choice. A fixed $N_r=10K$ for IS demonstrated to work well in our experiments, but this can be ensured via a preliminary run where we assess $N_r$ through monitoring the mixing of $\boldsymbol{\delta}$ and $\omega$. Additional simulation studies are included in the supplementary material investigating how $N_z$ increases with $d$ and $n$. 

In this study, 100 data sets are simulated with different configurations and sample sizes of the MultCOMP parameters. We adapt $N_z$ under the true parameter values and report the resulting mean and standard deviation. Results evidence that $N_z$ grows with the sample size but how they associate seems to depend on the parameter configuration. There is also a positive impact of increasing the data dimension, which is suggested by comparing configurations that take the same parameter values and vary $d$. In addition, higher COM-Poisson mean values relate to bigger $N_z$ since $\log z^{-1}$ will be higher in magnitude. For example, under $\lambda = 1$ and $\nu = 1.5$, the log reciprocal normalising constant is approximately $-0.89$, while this is around $-1.12$ when $\lambda = 1$ and $\nu = 0.7$. In these settings, the average $N_z$ is 3878 and 6672 respectively for $n=200$ which increases to 6672, 17137 when $n$ is 500. With these results, we are able to better understand on the effect of $N_z$ in the accuracy of the final log-likelihood estimator, which is crucial for a good performance of the GIMH. 

\subsection{Exchange algorithm}

The exchange algorithm by \cite{murray2006} provides a framework for conducting MCMC for a doubly-intractable problems relying on the ability to simulate exactly from the likelihood. It assumes that it is possible to write the model's likelihood function $p(\boldsymbol{X}|\boldsymbol\theta)$ as a product of a tractable unnormalised term $q(\boldsymbol{X}|\boldsymbol\theta)$ and the reciprocal normalising constant $\dfrac{1}{Z(\boldsymbol\theta)}$, which is intractable. By augmenting the target density with an auxiliary variable $\boldsymbol{X}' \sim p(\cdot|\boldsymbol\theta')$ where $\boldsymbol\theta'$ is a proposed state for $\boldsymbol\theta$, a cancellation of the normalising terms is achieved in the Metropolis-Hastings acceptance ratio. Algorithm \ref{exchange} gives the general formulation of an exchange algorithm where moves from the current state $\boldsymbol\theta^{(t)}$ to $\boldsymbol\theta'$ are proposed according to $h(\boldsymbol\theta'|\boldsymbol\theta^{(t)})$.

\begin{algorithm}

\small

\KwIn{Initial state $\boldsymbol\theta^{(t)}$}
 
Propose $\boldsymbol\theta' \sim h(\cdot|\boldsymbol\theta^{(t)})$;

Simulate $\boldsymbol{X}'$ from the likelihood at $\boldsymbol\theta'$, that is, $\boldsymbol{X}' \sim \dfrac{q(\cdot|\boldsymbol\theta')}{Z(\boldsymbol\theta')}$;

Accept $\boldsymbol\theta^{(t+1)} = \boldsymbol\theta'$ with probability
$$\pi = \min \left\{ 1, \dfrac{p(\boldsymbol\theta') q(\boldsymbol{X}|\boldsymbol\theta') q(\boldsymbol{X}'|\boldsymbol\theta^{(t)}) h(\boldsymbol\theta^{(t)}|\boldsymbol\theta') \bcancel{Z(\boldsymbol\theta)} \bcancel{Z(\boldsymbol\theta')}}{p(\boldsymbol\theta^{(t)}) q(\boldsymbol{X}|\boldsymbol\theta^{(t)}) q(\boldsymbol{X}'|\boldsymbol\theta')  h(\boldsymbol\theta'|\boldsymbol\theta^{(t)}) \bcancel{Z(\boldsymbol\theta')} \bcancel{Z(\boldsymbol\theta)}}\right\}.$$

\caption{Exchange algorithm}\label{exchange}
\end{algorithm}

The last step illustrates how the cancellation of normalising terms in achieved, leaving the acceptance ratio tractable. In this section, we investigate the application of the exchange algorithm for inferring on the multivariate COM-Poisson model parameters. Our motivation in developing this alternative approach is in avoiding the computation of $\widehat{z}^{-1}$ which can require a substantial number of auxiliary draws depending on $d$ and $n$.

However, in our context, an algorithm to simulate from the likelihood function exactly is not available. Instead, we formulate an noisy exchange algorithm where Step 2 of algorithm \ref{exchange} is done following Algorithm \ref{bcp_sampler}. Additionally, $q(\boldsymbol{X}|\boldsymbol\theta)$ is replaced with $\widehat{q}(\boldsymbol{X}|\boldsymbol\theta, \widehat{\boldsymbol{r}})$, an estimator of the MultCOMP unnormalised probability function. In analogy to \ref{exchange}, $Z(\boldsymbol{\theta}) \equiv Z(\boldsymbol{\lambda}, \boldsymbol{\nu})$ under our model. More specifically, $Z(\boldsymbol{\lambda}, \boldsymbol{\nu})$ is the product $\prod_{j=1}^d z_j^{-1}$ with $z_j^{-1} \equiv \dfrac{1}{Z(\lambda_j, \nu_j)}$. Under our formulation, there is inexactness in the cancellation of the normalising terms and approximation of $q({\bf X}|\boldsymbol\theta)$. Bearing this is mind, our goal is to compare the noisy exchange approach with GIMH under controlled scenarios. We highlight that there are extensions of the algorithm by \cite{murray2006} relaxing the perfect sampling requirement \citep{liang2016} but these still depend on an analytical calculation of ${q(\boldsymbol{X}|\boldsymbol\theta)}$. Algorithm \ref{exchange_bcp} describes single-site updates of the $\boldsymbol{\lambda}$ elements under our noisy exchange formulation. For increased computational efficiency, the ratio estimates are recycled throughout the iterations. This means that we propagate  $\boldsymbol{\widehat{r}}(\boldsymbol{\lambda}^{(t)}, \boldsymbol{\nu}^{(t)}, \omega^{(t)})$ and refresh $\widehat{r}_k(\lambda_k, \nu_k)$ whenever a new $\lambda_k$ or $\nu_k$ is accepted, or all elements when a new $\omega$ is visited. 

\begin{algorithm}

\small

\KwIn{$\boldsymbol\theta^{(t)}$, $\widehat{\bf r}^{(t)}$, $\boldsymbol{X}$}

\For{$j \rightarrow 1:d$}{

    Propose $\lambda_j' \sim \mbox{Log-Normal}(\log\lambda_j^{(t)}, \sigma_{\lambda_j})$ and set $\boldsymbol{\lambda}' = \boldsymbol{\lambda}^{(t)}[j] \leftarrow \lambda_j'$;
    
    Set $\boldsymbol\theta' = (\boldsymbol{\lambda}', \boldsymbol{\nu}^{(t)}, \boldsymbol{\delta}^{(t)}, \omega^{(t)})$;

     Estimate $\widehat{r}_j'$ and set $\widehat{\boldsymbol{r}}' \equiv \widehat{\boldsymbol{r}}^{(t)}[j] \leftarrow \widehat{r}_j'$
     
    Check if $\lambda_j'$ yields a valid joint probability function by ensuring that every component of $\boldsymbol\delta$ satisfies (\ref{mult_range}). If this is not achieved, return to the Step 2.
    
    Draw $\boldsymbol{X}' \sim \mbox{MultCOMP}(\boldsymbol\theta')$ (Sampler \ref{bcp_sampler});

     Calculate $$\pi = \min \left\{1, \dfrac{p(\lambda_1') h(\lambda_1^{(t)}|\lambda_1') \widehat{q}(\boldsymbol{X}| \boldsymbol\theta', \widehat{\boldsymbol{r}}' ) \widehat{q}(\boldsymbol{X}'|\boldsymbol\theta^{(t)}, \widehat{\boldsymbol{r}}^{(t)})}{p(\lambda_1^{(t)}) h(\lambda_1'|\lambda^{(t)})  \widehat{q}(\boldsymbol{X}| \boldsymbol\theta^{(t)}, \widehat{\boldsymbol{r}}^{(t)}) \widehat{q}(\boldsymbol{X}'| \boldsymbol\theta', \widehat{\boldsymbol{r}}') } \right\};$$

    With probability $\pi$ set $\lambda_1^{(t+1)} = \lambda_1'$ and $\widehat{\boldsymbol{r}}^{(t)} = \widehat{\boldsymbol{r}}'$.
}
\caption{Noisy Exchange Algorithm for the multivariate COM-Poisson model: $\boldsymbol{\lambda}$ update}\label{exchange_bcp}
\end{algorithm}

As before, $\boldsymbol{\nu}$ update is carried in a similar fashion to $\boldsymbol{\lambda}$ and that of $\omega$ requires all $\boldsymbol{\widehat{r}}'$ elements to be estimated. The final move for $\boldsymbol{\delta}$ is simplified due to the independence of the ratio on this parameter. Similarly to GIMH, we employ a truncated normal proposal in this step and the same previous prior specification.

\section{Simulation studies}\label{simulations}

\subsection{Comparing algorithms with bivariate data}\label{sim_algorithms}


This section is dedicated to comparing the proposed algorithms via simulation studies. The GIMH and noisy exchange approaches employing IS ratio estimators are applied to synthetic bivariate and trivariate COM-Poisson data. Under controlled settings, we can assess whether the inference is consistent with the true parameter values used to simulate the data and how GIMH and the noisy exchange results compare.

\begin{figure}
    \centering
    \includegraphics[width=.6\textwidth]{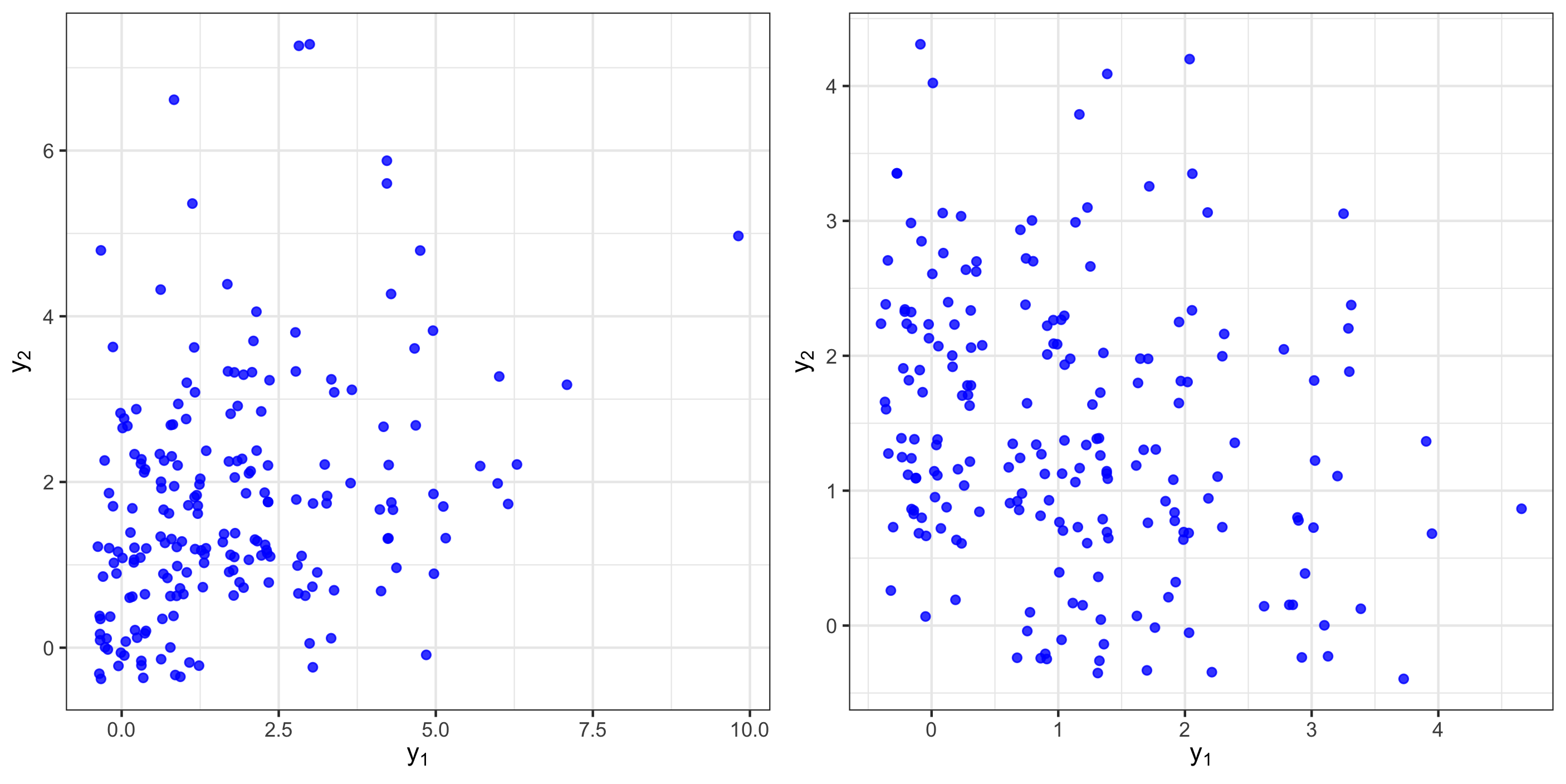}
    \caption{Scatter plots of data sets simulated under Configuration 1 (to the left) and Configuration 2 (to the right).}\label{datasets}
\end{figure}

We simulate data from the bivariate COM-Poisson distribution under two configurations following the steps in Algorithm \ref{bcp_sampler} with sample size $n=200$. In the first scenario, the two components are positively correlated and overdispersed. More specifically, we set $\lambda_1= 1$, $\lambda_2 = 1.5$, $\nu_1 = 0.4$, $\nu_2 = 0.8$, $\omega = 2$, and $\delta = 3$. The empirical means, variances, and Pearson's linear correlation for this synthetic data set are $(1.745, 1.745)$, $(2.915, 1.990)$, and $0.340$ respectively. Configuration 2 sets $\lambda_1= 1$, $\lambda_2 = 2$, $\nu_1 = 0.8$, $\nu_2 = 1.5$, $\omega = 1.5$, and $\delta = -2$, a scenario where the first and second components are overdispersed and equidispersed respectively and have a negative dependency. The correlation for this second data set is $-0.258$ and the marginal means and variances are respectively $(1.070, 1.420)$ and $(1.141, 0.968)$. Scatterplots with added jitter (small random noise) are displayed in Figure \ref{datasets} illustrating the two bivariate synthetic data sets. This is done via \verb|geom_jitter| from R package \verb|ggplot2| to improve visualization by avoiding that observations are plotted directly on top of each other.

For each data set, five parallel chains of the GIMH and noisy-exchange algorithms are run for 30K iterations with the first 10K  discarded as burn-in. Convergence of the multiple chains is assessed via the Brooks-Gelman-Rubin $\widehat{R}$ statistic \citep{bgr}. If $\widehat{R}$ is close to 1 for all model parameters, convergence is accepted and inference and posterior draws are combined. Prior distributions of the location parameters are set to $\mbox{Gamma}(2,2)$, while a $\mbox{Gamma}(1.5,2)$ is adopted for the parameters controlling the dispersion. Those of $\omega$ and $\delta$ are $\mbox{Gamma}(2,0.8)$ and $\mbox{Truncated-Normal}(0,5)$, respectively. Adaptation period for $N_z$ is employed before running GIMH targeting 1.2 likelihood standard. This returned $N_z \approx$ 18K for the first configuration (minimum 17708 and maximum 18318) and 15K for the second (minimum 14648 and maximum 15489). Furthermore, $N_r$ is fixed to 10K in both algorithms.

\begin{figure}
    \centering
    \includegraphics[width=0.8\textwidth]{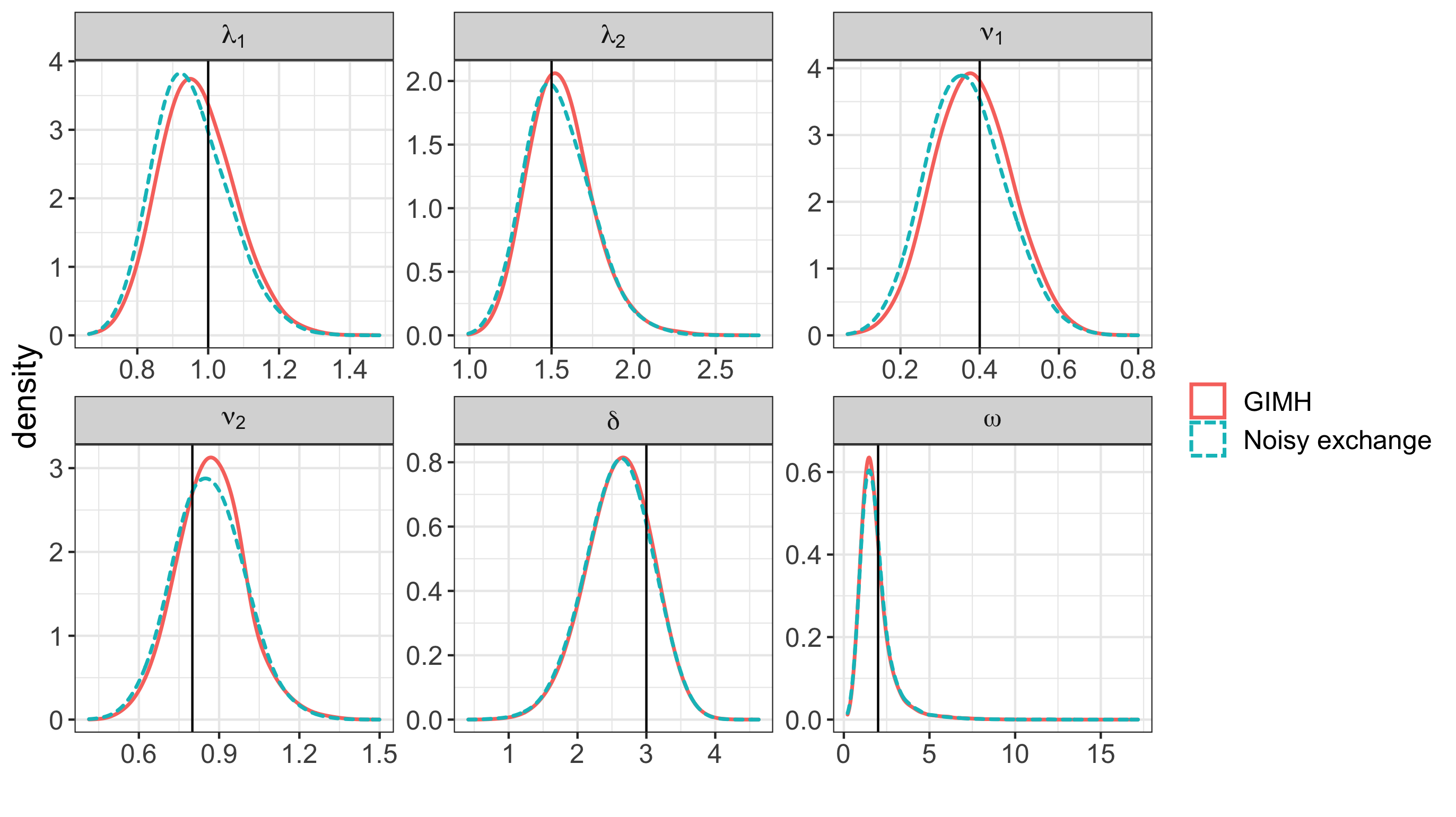}
    \caption{Posterior density plots of the bivariate COM-Poisson model fit to the synthetic data set of Configuration 1. Different line types correspond to the fit of the GIMH and noisy exchange algorithms and the vertical solid lines indicate the true parameter values used to generate the data.}\label{density_c1}
\end{figure}

The results due to the first configuration are illustrated in Figure \ref{density_c1}, where the posterior densities obtained under each method are displayed using different line types. These are produced from the 100K combined draws from the five parallel chains as we found $\widehat{R}$ values close to 1 for all parameters. Posterior density estimates from the two algorithms are very similar and mostly overlap. Among the model parameters, there is a higher uncertainty in the posterior distribution of $\omega$, which has a heavy right tail. This parameter plays the role of extending the correlation range supported by a given $\delta$ value. However, at some point a plateau is reached, in a way that there is no further increase in the correlation related to the increase in $\omega$. In our view, this explains the heavy right tail of this parameter's posterior distribution. Table \ref{table_c1} provides numerical summaries of the posterior distributions. As expected, the posterior mean is close to the parameter values used to generate the data, which are comprised by the 95\% percentile-based credible intervals in all cases.

\begin{table}
\small
\centering
\begin{tabular}{@{}clllclll@{}}
\toprule
\multirow{2}{*}{\textbf{Algorithm}} & \multicolumn{3}{c}{$\lambda_1$}                                                  & \multirow{2}{*}{\textbf{Algorithm}} & \multicolumn{3}{c}{\textbf{$\lambda_2$}}                                                    \\ \cmidrule(lr){2-4} \cmidrule(l){6-8} 
                                    & \multicolumn{1}{c}{Mean}  & \multicolumn{1}{c}{95\% CI} & \multicolumn{1}{c}{SD} &                                     & Mean                      & \multicolumn{1}{c}{95\% CI}         & \multicolumn{1}{c}{SD}    \\ \cmidrule(r){1-1} \cmidrule(lr){5-5}

Noisy Exchange                         & 0.953                     & $(0.801, 1.135)$             & 0.102                  & Noisy Exchange                & 1.545 & $(1.247, 1.895)$ & 0.200 \\
GIMH                           & 0.971                     & $(0.812, 1.152)$             & 0.104                  & GIMH                           & 1.559 & $(1.266, 1.911)$ & 0.199 \\
\multicolumn{1}{l}{}                & \multicolumn{3}{c}{\textbf{$\nu_1$}}                                             & \multicolumn{1}{l}{}                & \multicolumn{3}{c}{$\nu_2$}                                                              \\ \cmidrule(lr){2-4}    \cmidrule(l){6-8} 
Noisy Exchange                         & 0.366 & $(0.213, 0.535)$             & 0.097                 & Noisy Exchange                         & 0.863                     & $(0.653, 1.081)$                     & 0.131                     \\
 GIMH                           & 0.384                     & $(0.229, 0.548)$             & 0.097                 & GIMH                           & 0.871                     & $(0.666, 1.089)$                     & 0.128                   \\
\multicolumn{1}{l}{}                & \multicolumn{3}{c}{\textbf{$\delta$}}                                            & \multicolumn{1}{l}{}                & \multicolumn{3}{c}{\textbf{$\omega$}}                                                       \\ \cmidrule(lr){2-4}    \cmidrule(l){6-8} 
Noisy Exchange                         & 2.592                  & $(1.781, 3.334)$       & 0.471                  & Noisy Exchange                        & 1.862                     & $(0.804, 3.671)$                       & 1.008                     \\
 GIMH                           & 2.605                  & $(1.801, 3.346)$       & 0.470                  & GIMH                           & 1.846                     & $(0.799, 3.628)$                       & 0.999                     \\ \bottomrule
\end{tabular}\caption{Posterior mean, standard deviation (SD) and 95\% credible interval of the bivariate COM-Poisson model parameters obtained via the GIMH and noisy exchange algorithms for the synthetic data set under Configuration 1.}\label{table_c1}
\end{table}

\begin{figure}
    \centering
    \includegraphics[width=0.8\textwidth]{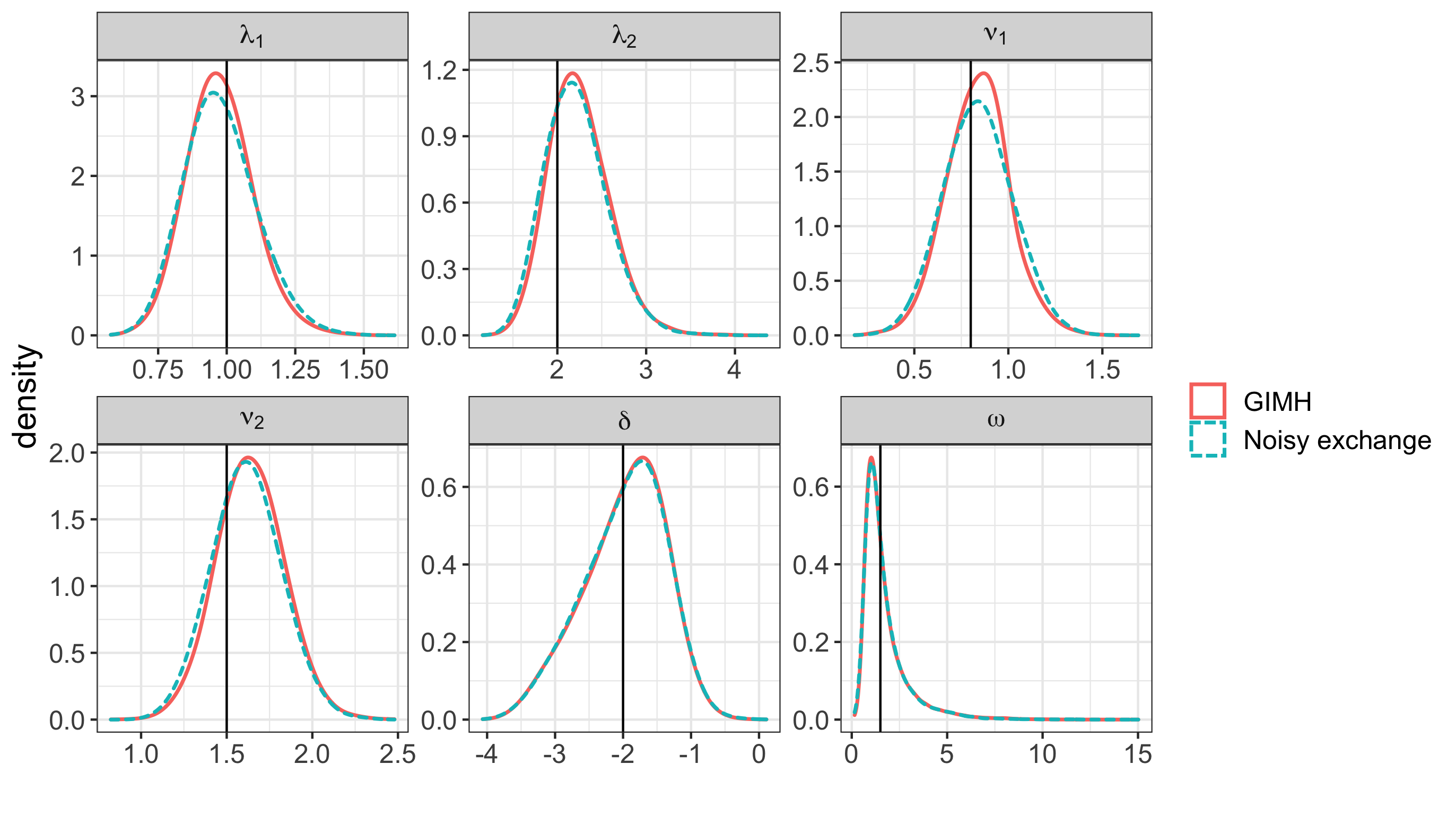}
  \caption{Posterior density plots of the bivariate COM-Poisson model fit to the synthetic data set under Configuration 2. Different line types correspond to the fit of the GIMH and noisy exchange algorithms and the vertical solid lines indicate the true parameter values used to generate the data.}\label{density_c2}
\end{figure}

The model fits for the second simulated data set are presented in Figure \ref{density_c2} with numerical summaries given in Table \ref{table_c2}. As before, there is agreement between the MCMC methods as evidenced by the proximity of posterior curves obtained from each algorithm. The numerical summaries also show that the parameters used to generate the data are very likely under the posterior model, all covered by the 95\% credible intervals.

\begin{table}
\small\centering
\begin{tabular}{@{}clllclll@{}}
\toprule
\multirow{2}{*}{\textbf{Algorithm}} & \multicolumn{3}{c}{$\lambda_1$}                                                  & \multirow{2}{*}{\textbf{Algorithm}} & \multicolumn{3}{c}{\textbf{$\lambda_2$}}                                                    \\ \cmidrule(lr){2-4} \cmidrule(l){6-8} 
                                    & \multicolumn{1}{c}{Mean}  & \multicolumn{1}{c}{95\% CI} & \multicolumn{1}{c}{SD} &                                     & Mean                      & \multicolumn{1}{c}{95\% CI}         & \multicolumn{1}{c}{SD}    \\ \cmidrule(r){1-1} \cmidrule(lr){5-5}
Noisy Exchange                        & 0.977                     & $(0.781, 1.203)$             & 0.129                  & Noisy Exchange                & 2.213 & $(1.702, 2.810)$ &0.341 \\
 GIMH                           & 0.976                     & $(0.788, 1.182)$             & 0.121                  &  GIMH                           & 2.244 & $(1.740, 2.829)$ & 0.338 \\
\multicolumn{1}{l}{}                & \multicolumn{3}{c}{\textbf{$\nu_1$}}                                             & \multicolumn{1}{l}{}                & \multicolumn{3}{c}{$\nu_2$}                                                              \\ \cmidrule(lr){2-4}    \cmidrule(l){6-8} 

Noisy Exchange                         & 0.841 & $(0.550, 1.141)$             & 0.179                  & Noisy Exchange                         & 1.617                     & $(1.295, 1.949)$                     & 0.198                     \\
 GIMH                           & 0.839                     & $(0.570, 1.116)$             & 0.166                  &  GIMH                           & 1.639                     & $(1.320, 1.963)$                     & 0.197                     \\
\multicolumn{1}{l}{}                & \multicolumn{3}{c}{\textbf{$\delta$}}                                            & \multicolumn{1}{l}{}                & \multicolumn{3}{c}{\textbf{$\omega$}}                                                       \\ \cmidrule(lr){2-4}    \cmidrule(l){6-8} 

Noisy Exchange                          & $-1.981$                  & $(-3.066, -1.126)$       & 0.591                  & Noisy Exchange                        & 1.710                     & $(0.702, 4.118)$                       & 1.190                     \\
 GIMH                           & $-1.978$                  & $(-3.084, -1.122)$       & 0.594                  & GIMH                           & 1.728                     & $(0.699, 4.168)$                       & 1.220                     \\ \bottomrule
\end{tabular}\caption{Posterior mean, standard deviation (SD) and 95\% credible interval of the bivariate COM-Poisson model parameters obtained via the GIMH and noisy exchange algorithms for the synthetic data set under Configuration 2.}\label{table_c2}
\end{table}

\subsection{Trivariate data experiment}

Given the agreement between the algorithms compared in Subsection \ref{sim_algorithms}, we now present a tri-dimensional data example for which the fastest inference strategy will be employed. Given that $N_z$ depends on $d$ and $n$, the preferred algorithm can vary depending on the data set. In this experiment, where $d=3$ and $n=500$, preliminary runs of both algorithms indicate GIMH to be preferred.

The characteristics of the trivariate COM-Poisson simulated data set are as follows. The location and dispersion parameter vectors are respectively $\boldsymbol{\lambda} = (1.5, 1, 0.5)^\top$ and $\boldsymbol{\nu} = (1, 0.5, 0.8)^\top$ (the first component is equidispersed and the others are overdispersed). The parameters controlling the correlation are $\omega = 3$, $\delta_{12} = 3.5$, $\delta_{13}= -2.5$, $\delta_{23} = -3$, in a way that there are positively and negatively dependent pairs. The Pearson's correlation is $0.167$ for the pair $(X_1,X_2)$, $-0.165$ for $(X_1,X_3)$, and $-0.191$ is due to $(X_2,X_3)$.

As before, five parallel chains of GIMH from random starting points are run for a total of 30K iterations. Adaptation of $N_z$ resulted in $N_z$ around 15K for all chains (minimum 14525 and maximum 157891), also searching for a 1.2 likelihood standard deviation. Table \ref{triv_simulation} displays summary statistics of the parameters posterior distributions and density plots can be found in the Supplementary Material. The results are consistent with the parameters used to generate the data, which demonstrate to be likely under the parameters posterior distributions.

\begin{table}
\small
\centering
\begin{tabular}{@{}lccccllcccc@{}}
\toprule
\textbf{}         & \textbf{Mean} & \textbf{SD} & \textbf{Q5} & \textbf{Q95} & \textbf{} & \textbf{}           & \textbf{Mean} & \textbf{SD} & \textbf{Q5} & \textbf{Q95} \\ \midrule
$\lambda_1 = 1.5$ & 1.448         & 0.127       & 1.254       & 1.673        &           & $\omega =3$         & 3.421         & 1.601       & 1.682       & 6.588        \\
$\lambda_2 =1$    & 0.926         & 0.066       & 0.824       & 1.040        &           & $\delta_{12} = 3.5$ & 3.203         & 0.551       & 2.212       & 4.001        \\
$\lambda_3 =0.5$  & 0.529         & 0.045       & 0.458       & 0.606        &           & $\delta_{13}= -2.5$ & $-$2.719        & 0.495       & $-$3.497      & $-$1.867       \\
$\nu_1 =1$        & 0.970         & 0.101       & 0.808       & 1.144        &           & $\delta_{23} =3$  & $-$2.898        & 0.525       & $-$3.722      & $-$1.992       \\
$\nu_2 = 0.5$     & 0.445         & 0.073       & 0.326       & 0.567        &           &                     &               &             &             &              \\
$\nu_3= 0.8$      & 0.734         & 0.164       & 0.470       & 1.001        &           &                     &               &             &             &              \\ \bottomrule
\end{tabular}
\caption{Mean, standard deviation (SD) and 5\%, 95\% quantiles (Q5, Q95) of the trivariate COM-Poisson posterior model parameters. Results are due to the GIMH algorithm which demonstrated to be the fastest option for this simulated data set. }\label{triv_simulation}
\end{table}

\section{Premier League data analysis}\label{data_application}

We here illustrate the usefulness of the proposed multivariate COM-Poisson model in modelling real-life correlated count data. A novel data analysis that we focus on concerns the number of goals scored by the home and away team at Premier League matches. Our main goal is to assess the effect of the absence of crowds during the COVID-19 pandemic on the well-known home team advantage. Several papers have modelled soccer data demonstrating a positive effect of playing at home such as \cite{dixoncoles1997}, \cite{karlis2000}, and \cite{karlis2003}. Very recently, there is also a great interest in evaluating whether this has changed due to the absence of crowds at games during the pandemic; for instance, see \cite{tilp2020} and \cite{mccarrick2020}. The data we consider here consists of the outcome of games from the 2018-2019, 2019-2020 and 2020-2021 seasons. Each season invovles 380 matches. The Premier League resumed on June 17$^{th}, 2020$ after a break from March $9^{th}, 2020$ to the end of the $2020-2021$ season. During this period, due to the public health restrictions, no crowds were present at these matches. This results in a total of 1140 observations, 668 pre-pandemic games with crowds and 472 matches during the pandemic with no crowds present. A preliminary analysis is given in Table \ref{descriptive_premier} where the proportion of home draws, losses and wins are compared for these two distinct time periods, pre- and during-pandemic. A reduction in the proportion of home team wins is observed during the pandemic when crowds are present with an associated increase in losses, supporting the hypothesis that the advantage of playing at home is decreased in the pandemic matches.

\begin{table}[H]
\small
\centering
\begin{tabular}{@{}lccc@{}}
\toprule
\textbf{}    & \textbf{Draws} & \textbf{Losses} & \textbf{Wins} \\ \midrule
Pre-pandemic & 21.4          & 32.2           & 46.4         \\
During-pandemic     & 21.8          & 38.6           & 39.6         \\ \bottomrule
\end{tabular}
\caption{Proportion of Premier League home team draws, losses and wins pre and during the COVID-19 pandemic from 2018 to 2021.}\label{descriptive_premier}
\end{table}

Over the entire study period, the marginal mean and variance of the number of goals for the home team are $(1.48, 1.68)$, respectively and $(1.27, 1.47)$, respectively for the away team, while the empirical correlation between goals scored at home and away is $-0.162$. Calculating these statistics pre- and during-pandemic yields a marginal mean and variance of $(1.54, 1.59)$ pre-pandemic and $(1.39, 1.80)$ during-pandemic for the home team. Similarly, both statistics for the number of goals scored by the way team  are respectively $(1.24, 1.42)$ pre-pandemic and $(1.31, 1.55)$ during-pandemic. The overdispersion and negative correlation in the data motivate us to consider the MultCOMP model which is able to accommodate  both of these features of the data. Further, we will adopt a regression structure on the count of home team goals given that this seems to decrease for matches during the pandemic. We take $X_{1i}$ and $X_{2i}$ to be the number of goals scored by the home and away team, respectively, with $i =1,\cdots, 1140$ denoting an index to each Premier League match. The following assumptions are made in our data analysis. The pair $(X_{1i},X_{2i}$) is a multivariate observation from a MultCOMP distribution with match-specific location parameters. That is, $(X_{1i},X_{2i})\sim\mbox{MultCOMP}(\boldsymbol\lambda_i,\boldsymbol\nu,\boldsymbol\delta,\omega)$ with $\boldsymbol\lambda_i=(\lambda_{1i},\lambda_{2i})^\top$. Further, we assume independence among matches and the following regression structure on $\boldsymbol\lambda_i$:
\begin{eqnarray}\label{regression_reduced}
\begin{cases}
\log\lambda_{1i} = \gamma_0 + \gamma_1\texttt{Home}_i + \gamma_2 \texttt{Pandemic}_i,\\ 
\log\lambda_{2i} = \gamma_0, 
\end{cases}
\end{eqnarray}
for $i=1,\ldots,1140$, where \texttt{Home} ($=1$ for home team and $=0$ otherwise) and \texttt{Pandemic} ($=1$ for matches realized during the pandemic and $=0$ otherwise) are indicator covariates. Following \cite{karlis2000, karlis2003} and other references in the field, we adopt a common intercept $\gamma_0$ for the competing teams. Since the home team goals are assigned to the first component, the first indicator is always one ($\texttt{Home}_i = 1$ $\forall i$) and represents a deviation from the overall average number of goals scored by a team in a Premier League match. This is explicit in the regression structure for clarity and implies that $\gamma_1$ measures the home team main effect during the pre-pandemic period. When $\verb|Pandemic|=1$,  the difference between home and away teams becomes $\gamma_1 + \gamma_2$ so we can interpret $\gamma_2$ as the parameter measuring the pandemic effect on the home team advantage. We shall denote the elicited model $\mathcal{M}_{BCOMP}$.
 
The Bayesian inferential procedures introduced in the paper holds for the iid case and are be easily adapted for the regression  analysis involving categorical covariates, with the developed methodology being applied to each regressor level. Inference is carried with 100K draws from the model posterior distributions resulting from the combination of five parallel chains of the GIMH algorithm. Adaptation of $N_z$, as described in Subsection \ref{mixing_gimh}, indicates that on average 170K auxiliary draws are necessary to ensure good mixing of the GIMH algorithm. Summaries of the posterior model parameter distributions are shown in Table \ref{post_premier} and respective density plots can be found in the Supplementary Material.

\begin{figure}
	\centering
\begin{minipage}[b]{.4\textwidth}
\begin{table}[H]
\small
\centering
\begin{tabular}{@{}lccccc@{}}
\toprule
           & \textbf{Mean} & \textbf{SD} & \textbf{Q5} & \textbf{Q50} & \textbf{Q95} \\ \midrule
$\gamma_0$ & 0.061         & 0.053       & $-0.023$    & 0.061        & 0.151        \\
$\gamma_1$ & 0.219         & 0.081       & 0.083       & 0.222        & 0.350        \\
$\gamma_2$ & $-0.087$      & 0.047       & $-0.164$    & $-0.086$       & $-0.010$     \\
$\nu_1$    & 0.818         & 0.063       & 0.714       & 0.819        & 0.920        \\
$\nu_2$    & 0.756         & 0.065       & 0.650       & 0.755        & 0.865        \\
$\delta$   & $-1.767$      & 0.355       & $-2.281$    & $-1.805$       & $-1.122$     \\ 
$\omega$ & 0.453      & 0.098       & 0.331    & 0.436       & 0.632   \\   
\bottomrule
\end{tabular}
\caption{Mean, standard deviation (SD) and 5\%, 50\% and 95\% quantiles (Q5, Q50, Q95) of the bivariate COM-Poisson posterior model parameters fitted to the Premier League data set. $\gamma_2$ measures the COVID-19 pandemic effect on the home team advantage $\gamma_1$.}\label{post_premier}
\end{table}
	\end{minipage}\hfill
	\begin{minipage}[b]{.5\textwidth}
 \includegraphics[width=0.9\textwidth]{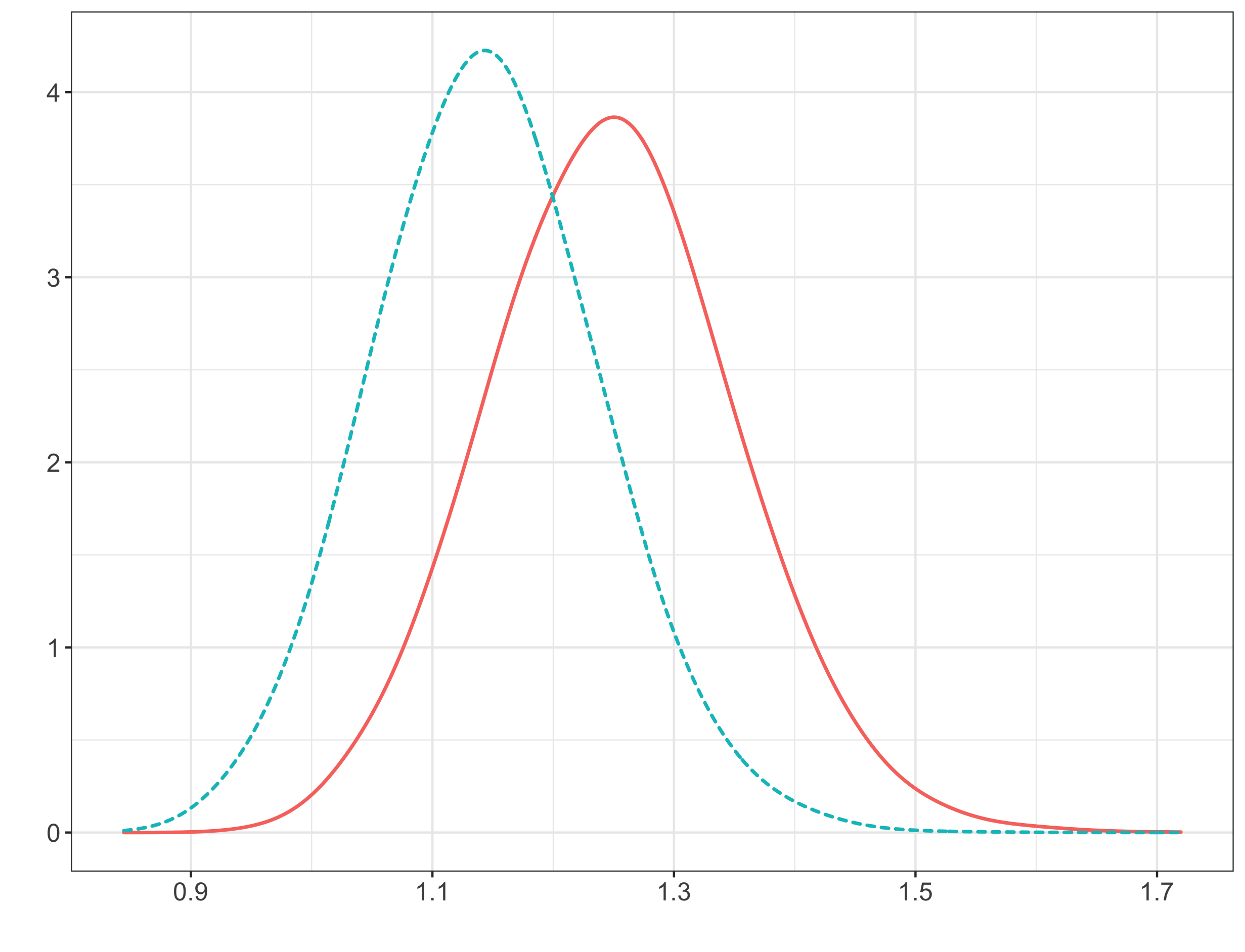}
 \caption{Posterior density of home team advantage in the Premier League before ($\exp\{\gamma_1\}$, solid line) and during ($\exp\{ \gamma_1 + \gamma_2\}$, dashed line) the COVID-19 pandemic.}\label{exp_params}
\vspace{2cm}
	\end{minipage}
\end{figure}

The posterior distributions of the MultCOMP model parameters are in accordance with our preliminary analysis. The negative dependency between $X_{1i}$ and $X_{2i}$ is captured by the proposed model as determined by the sign of $\delta$. Moreover, the 99\% credible interval of this parameter ($-2.472$ - $-0.741$) does not contain zero, the independence case. It is also seen that the marginal mean-variance relationships are well modelled with both $\nu_1$ and $\nu_2$ below one, which captures the data overdispersion.

In the absence of pandemic, the home goals surplus is measured by $\gamma_1$ which we can interpret on a multiplicative scale by taking $\exp(\gamma_1)$. The posterior expectation of this quantity is $1.249$, with associated 95\% credible interval of $(1.055 - 1.455)$. Hence, playing at home when crowds are present increases the goals scored, on average, by around 25\%. Without the public, the home effect on the number of goals is $\exp \{\gamma_1 + \gamma_2\}$ which has a posterior expectation of $1.146$ and a 95\% credible interval of $(0.968 - 1.336)$. The shift in home team advantage is illustrated in Figure \ref{exp_params}, where the posterior density of $\exp\{\gamma_1\}$ and $\exp\{\gamma_1 + \gamma_2\}$ are displayed. Since $P(\gamma_2 <0) = 0.97$, we conclude that there is a high posterior probability that the home team advantage, here measured by the home goals surplus, has decreased with the absence of crowds in Premier League matches.


Alternative models were fitted and compared to $\mathcal{M}_{BCOMP}$ via the Pareto-Smoothed Importance Sampling Leave-One-Out (PSIS-LOO) criterion. While the Bayes factor or model evidence is often used to quantify the support for competing models, we adopt PSIS-LOO to overcome the difficulty in computing the MultCOMP model evidence. Proposed by \cite{vehtari2017}, PSIS-LOO is a fully Bayesian model information criterion that is based on the idea of fitting the model without each individual data point and evaluating the likelihood of the left out point under the posterior distribution. Naturally, this would involve computing the model posterior $n$ times, which becomes computationally expensive for larger data sets. The PSIS method avoids model refit by approximating such posterior distributions through a combination of importance sampling and a Pareto distribution fit to the upper tail of importance weights. We refer to the R package \verb|loo| \citep{loo} to compute this criterion for each competing model, where the one with minimum PSIS-LOO is expected to have the best predictive performance. 

A more complex regression structure was considered for the MultCOMP parameters in the model which we denote as $\mathcal{M}_{BCOMP, Full}$. In this version, the pandemic effect is component-specific and is also included in the model's dependency structure. This is done by setting $\log\lambda_{2i} = \gamma_0 + \gamma_3 \texttt{Pandemic}_i$ and $\log \gamma = \alpha_0 + \alpha_1 \texttt{Pandemic}_i$ where $\alpha_1$ is the parameter quantifying the pandemic effect in the correlation. A bivariate COM-Poisson model without the pandemic covariate ($\mathcal{M}_{BCOMP, 0}$) was also fitted to the data as well as the bivariate Poisson special case ($\nu_1 = \nu_2 = 1$) with the regression structure (\ref{regression_reduced}). The former is denoted by $\mathcal{M}_{P}$, a tractable model that we fit with a Gibbs sampler.

The expected PSIS-LOO for $\mathcal{M}_{BCOMP}$ is 6917.2 (32.6) with estimated standard deviation in parenthesis. While this is 6979.8 (63.8) for $\mathcal{M}_{BCOMP, Full}$, 6923.1 (64.0) for $\mathcal{M}_{BCOMP, 0}$ and finally 6938.4 (71.6) for $\mathcal{M}_{P}$. Although the smallest expected value is due to $\mathcal{M}_{BCOMP}$, we cannot decisively choose between models according to their predictive performances given that the differences in PSIS-LOO are not high in comparison to their variability. We proceed by examining the posterior distribution of nested model fits from which the following observations are made. The $\mathcal{M}_{BCOMP}$ model would reduce to $\mathcal{M}_{P}$ if equidispersion was a reasonable assumption for the marginal count distributions. From Table \ref{post_premier}, we have that $P(\nu_1 < 1)$ and $P(\nu_2 < 1)$ are close to one under $\mathcal{M}_{BCOMP}$ which motivates us to choose $\mathcal{M}_{BCOMP}$ over its Poisson special case. Regarding $\mathcal{M}_{BCOMP, Full}$, it does not seem to be worthwhile considering this more complex model given that the posterior distributions of $\gamma_2$ and $\gamma_3$ are highly similar and that $\alpha_1$ is concentrated at zero as shown in the Supplementary Material.

Finally, a comparison between $\mathcal{M}_{BCOMP}$ and the baseline model $\mathcal{M}_{BCOMP, 0}$ is carried out by drawing from their posterior predictive distributions. In this analysis, 100K data sets pre and during pandemic are simulated from each model and summarised according to relevant statistics. Our interest is in evaluating how well the alternative models capture characteristics of the original data, which is done by comparing the posterior predictive distribution of the statistics to their observed values.

\begin{figure}
    \centering
    \includegraphics[width=0.5\linewidth]{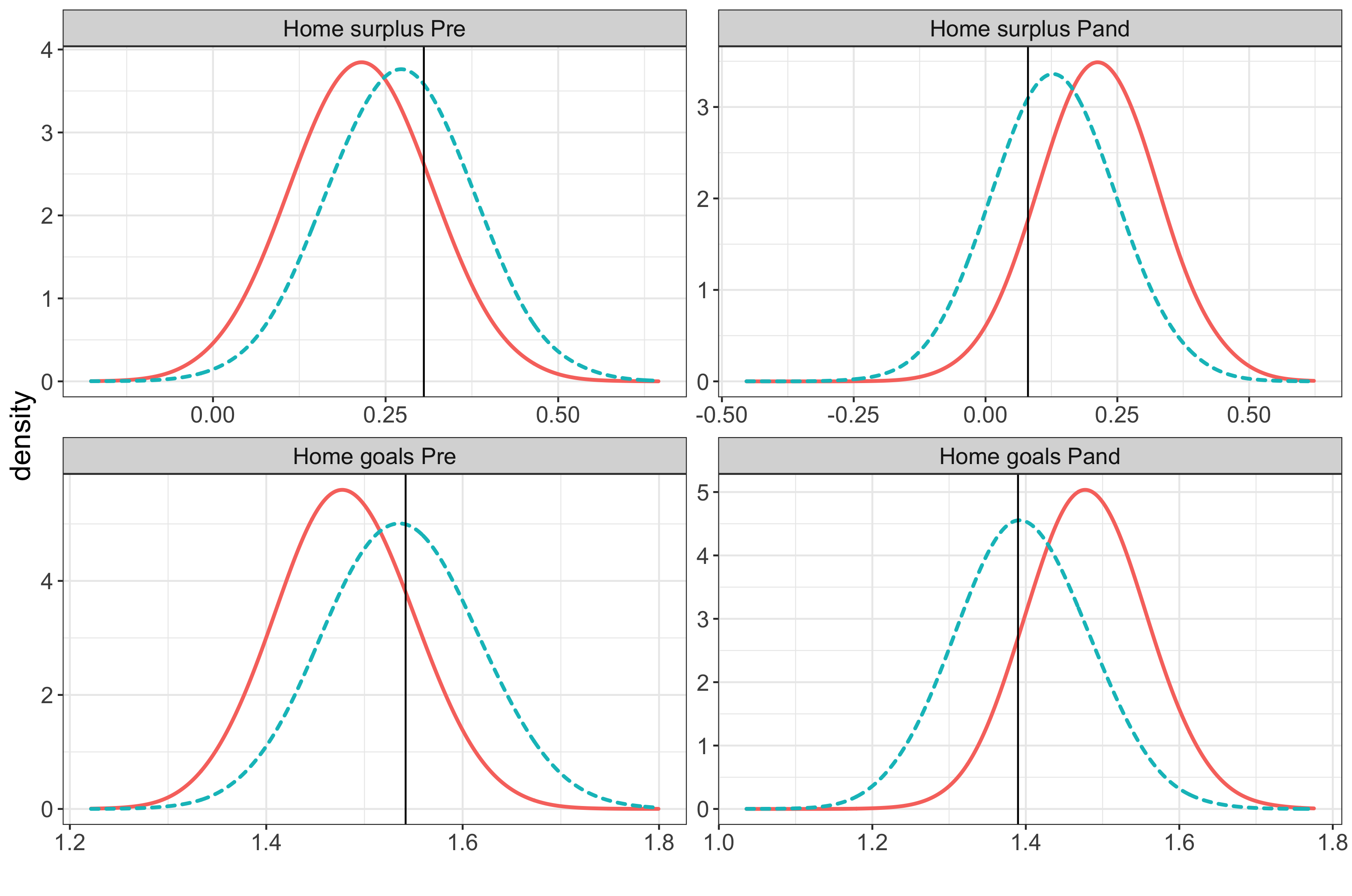}
    \caption{Posterior predictive density plots of the average number of goals and goals surplus of the home team at Premier League matches pre and during the COVID-19 pandemic. Dashed lines are due to the data sets replicated under $\mathcal{M}_{BCOMP}$ while those in solid are due to $\mathcal{M}_{BCOMP,0}$. Vertical lines indicate the observed values of each statistics.}\label{modelcheck_premier}
\end{figure}

Let $(\boldsymbol{X}_{1}, \boldsymbol{X}_{2})^{k,pre}$, $(\boldsymbol{X}_{1}, \boldsymbol{X}_{2})^{k,pand}$ denote the $k^{th}$ pre and during pandemic simulated data sets either from $\mathcal{M}_{BCOMP}$ or $\mathcal{M}_{BCOMP, 0}$, for $k=1, \cdots, 100K$. In keeping with the Premier League data characteristics and given that there is no pandemic effect under $\mathcal{M}_{BCOMP, 0}$, the difference between the two sets when simulating from this model is simply the sample size. If the data is replicated according to $\mathcal{M}_{BCOMP}$, $(\boldsymbol{X}_{1}, \boldsymbol{X}_{2})^{k}_{pand}$ includes the effect of $\gamma_2$ in $\log \lambda_1$. Given our interest in the shift of home advantage, we record the average number of goals scored by the home team pre and during the pandemic \texttt{Home goals}$^{pre}$, \texttt{Home goals}$^{pand} \equiv$ $\sum_{i=1}^{668} X^{k, pre}_{i,1}/668, \quad \sum_{i=1}^{472} X^{k, pand}_{i,1}/472$ and the average surplus
\texttt{Home surplus}$^{pre}$, \texttt{Home surplus}$^{pand} \equiv$ $\sum_{i=1}^{668} (X^{k, pre}_{i,1} - X^{k, pre}_{i,2})/668, \quad \sum_{i=1}^{472} (X^{k, pand}_{i,1} - X^{k, pand}_{i,2})/472$. The density plots in Figure \ref{modelcheck_premier} illustrate the results from the proposed posterior predictive analysis. Distributions drawn with dashed lines are those due to $\mathcal{M}_{BCOMP}$ which are closer to the observed values (vertical lines) than the null model (in solid). As per the bottom-right window, the number of home team goals expected by the null model is quite higher than what is observed during the pandemic, a decrease that is well explained by the lack of public in these matches. 

\section{Concluding remarks}\label{conclusions}

A $d$-dimensional COM-Poisson model was proposed to deal with multivariate correlated counts, which accommodates both positive and negative dependency, underdispersion, overdispersion, and equidispersion. To achieve this aim, we proposed a modified Sarmanov method which can be applied for other cases. Advantages over of model over existing bivariate COM-Poisson distributions, such the model by \cite{seletal2016}, were addressed.

Careful and detailed Bayesian inferential procedures were developed to treat the doubly-intractable likelihood challenge due to our model construction. Different inferential strategies for the doubly-intractable posterior distribution of the multivariate COM-Poisson model were proposed in Section \ref{inference}. The first option is the GIMH, a pseudo-marginal approach that relies on an unbiased likelihood estimator. Since the number of auxiliary draws required for the $\boldsymbol{z}^{-1}$ estimators involved in this approximation increases with $d$ and $n$, an alternative noisy exchange algorithm was proposed. Although this option is inexact, its advantage is in avoiding the computation of $\boldsymbol{z}^{-1}$.

Simulation studies were conducted to investigate the two inference strategies. Results from artificial bivariate data sets showed that the GIMH and noisy exchange algorithm produced very similar results, with negligible difference among them. Moreover, the posterior means were close to the parameter values used to generate the data, all comprised by the 95\% credible intervals. This demonstrates that both options provide sensible inference for the proposed model and computational speed can guide the choice of algorithm to be used in each application. Finally, a trivariate example displaying positively and negatively related components was included to illustrate the $d>2$ case.

An empirical illustration to investigate the impact of the COVID-19 on the Premier League was presented based on the methodologies developed in this paper. We fitted a bivariate COM-Poisson regression model to the goals scored by the home and away teams in the Premier League from 2018 to 2021, also considering the effect of no crowds during the COVID pandemic. Our inferential analysis has showed a potential decrease in the number of goals scored by the home team during the pandemic compared to number of goals scored pre-pandemic.

We also analysed a shunters accident data \citep{arbous1951} in the Supplementary Material, which is a well-known example that was used to illustrate numerous count data models in the literature; for instance, see \cite{aitchison1989}, \cite{famoye1995}, \cite{seletal2016}, and \cite{jones2019}. Here we conduct a full Bayesian analysis employing our model, its Poisson special case (BP-S), the trivariate reduction bivariate Poisson (BP-T), and the bivariate negative binomial (BNB) distribution by \cite{marshall1990}. Results reported in the Supplementary Material showed that the proposed model is preferred to the Poisson alternatives and is at least competitive with respect to the BNB model.

\section*{Acknowledgements}
L.S.C. Piancastelli and N. Friel wish to acknowledge the financial support of Science Foundation Ireland under Grant Numbers 18/CRT/6049 and 12/RC/2289 P2. 
W. Barreto-Souza and H. Ombao would like to acknowledge support by KAUST Research Fund and NIH 1R01EB028753-01.

\bibliographystyle{chicago}
\scriptsize
\bibliography{refs}

\end{document}